\pdfoutput=1
\documentclass[aps, prd, floats, floatfix, superscriptaddress, onecolumn,PRD,nofootinbib,amsmath,amssymb]{revtex4}

\usepackage[T1]{fontenc}
\usepackage[utf8]{inputenc}
\usepackage{lmodern}
\usepackage{lipsum}
\usepackage{enumerate}

\usepackage{MnSymbol}
\usepackage{amssymb}
\usepackage{amsfonts}
\usepackage{eufrak}

\usepackage[dvipsnames, usenames]{xcolor}
\usepackage{graphicx}
\usepackage{xspace}
\usepackage{amssymb}
\usepackage[normalem]{ulem} 
\usepackage{bm} 
\usepackage{appendix}

\definecolor{linkcolor}{rgb}{0.0,0.3,0.5}
\usepackage[hypertexnames=false, unicode, colorlinks=true, linkcolor=linkcolor,
citecolor=linkcolor, filecolor=linkcolor,urlcolor=linkcolor,
pdfusetitle]{hyperref}

\newcommand*{\fnl} {f_{\rm NL}}

\definecolor{darkred}{RGB}{175,0,0}
\definecolor{darkblue}{RGB}{14,0,185}

\begin{document}

	\title{3D-Radial galaxy correlation function}
	
	\newcommand\francescohome{
		\affiliation{Dipartimento di Fisica Galileo Galilei, Universit\` a di Padova, I-35131 Padova, Italy}
		\affiliation{INFN Sezione di Padova, I-35131 Padova, Italy}
	}

	\newcommand\alvisehome{
		\affiliation{Dipartimento di Fisica Galileo Galilei, Universit\` a di Padova, I-35131 Padova, Italy}
		\affiliation{INFN Sezione di Padova, I-35131 Padova, Italy}
		\affiliation{INAF-Osservatorio Astronomico di Padova, Italy}
	}

	\author{Francesco Spezzati}
	\francescohome
	
	\author{Alvise Raccanelli}
	\alvisehome

	\begin{abstract}
		Tests of cosmological models via measurements of galaxy correlations will require increasing modeling accuracy, given the high precision of measurements promised by forthcoming galaxy surveys.
		In this work we investigate the biases introduced in parameter estimation when using different approximations in the modeling of the galaxy two point correlation function.
		We study this for two example surveys, with different binning strategies, for measurements of the Primordial non-Gaussianity parameter $\fnl$ and the growth rate of structures $\gamma$.
		We then investigate the same issue for the nDGP model, to see if results will change for a different cosmological model.
		Our results show that failing to properly account for radial and angular separation between galaxies will induce a considerable shift in parameters best fit estimates, the bias being larger for thicker redshift bins. When accounting for radial evolution within the bins by integrating over $z$, such shifts are reduced but still present.
		We then introduce a new hybrid model, which we call 3D radial, where we neglect the purely wide angle terms, but include a proper 3D modeling of the system by including radial modes.
		Using this model, we show that biases are greatly reduced, making it an accurate formalism to be safely used for forthcoming galaxy surveys. This moreover confirms other recent findings on the importance of including radial modes to accurately model the galaxy correlation function.
	\end{abstract}
	
	\maketitle 
	
	\section{Introduction}
	\label{sec:intro}
	In the coming decade, the study of galaxy clustering promises to reach its golden age. It is by now well established that massive investments on new instruments, both from the ground and from space, will deliver data with unprecedented precision. Example of this are the current Euclid mission~\cite{Euclid:2024yrr,EUCLID:2011zbd,Euclid:2019clj} and DESI~\cite{DESI:2016fyo,DESI:2024mwx},  the future SPHEREx All Sky Survey~\cite{Dore:2014cca}, Vera Rubin Observatory~\cite{LSSTDarkEnergyScience:2012kar}, PFS~\cite{PFSTeam:2012fqu} and Nancy Roman Space Telescope~\cite{Eifler:2020vvg,Wenzl:2021rrq} and the proposed MegaMapper~\cite{Schlegel:2022vrv}, ATLAS~\cite{Wang2018} and SIRMOS~\cite{XXX}. However, on the theoretical side, there needs to be a comparable effort in providing the maximum level of accuracy possible (both on the modeling of the observables and the physical interpretation), especially considering that different galaxy surveys will focus on different regimes and parts of the observational parameter space.
	Theoretical modeling efforts in this direction started a while ago, covering various aspects, from the treatment of systematic errors~\cite{Brieden:2020upf,Kitching:2016hvn,Valcin:2021jcg,Norena:2011sh} to the modeling of small-scale physics~\cite{Bernardeau:2001qr,Matarrese:2007wc,Pietroni:2008jx,Matsubara:2007wj,Matsubara:2008wx,Takahashi:2012em,Chen:2020fxs,Reid:2011ar,Vlah:2018ygt,Vlah:2015zda,Vlah:2015sea,DAmico:2021rdb,Fasiello:2022lff,Senatore:2014vja,Perko:2016puo,Porto:2013qua,Carrasco:2012cv,Baumann:2010tm,Wang:2022itv,Philcox:2019hdi,Taruya:2010mx,Scoccimarro:2004tg,delaTorre:2012dg}, to more purely mathematical and geometrical modeling, focusing mostly on larger scales~\cite{Szalay:1997cc,Bertacca:2012tp,Bertacca:2017dzm,Bertacca:2019wyg,Raccanelli:2010hk,Raccanelli:2012gt,Raccanelli:2013dza,Raccanelli:2013gja,Raccanelli:2015vla,Raccanelli:2016avd,Raccanelli:2023fle,Raccanelli:2023zkj,Elkhashab:2021lsk,Papai:2008bd,Matsubara:1999du,Zaroubi:1993qt,DiDio:2013sea,Bonvin:2011bg,Castorina:2017inr,Castorina:2019hyr,Jeong:2011as,Yoo:2014kpa,Challinor:2011bk,Bonvin:2005ps,Yoo:2010ni, Yoo:2012se, Montanari:2015rga}.
	In this paper we focus on the latter aspect; in particular, when the precision guaranteed by the instrument is high, the theoretical modeling for the statistical quantities we measure must be able to provide a comparable accuracy.
	
	One question becomes if this means that it will be necessary to
	encapsulate the full 3-dimensionality of the system formed by the observer and the sources, without employing approximations that compress the full information extracted from the observed position of each source in a given cosmic volume into an approximate and less informative one. Such compression, in fact, can lead to a decrease in the accuracy in the measurements of cosmological parameters carried out with future galaxy survey, leading even to a bias in our understanding of the underlying physics.
	
	In this work we focus on the galaxy two point correlation function at large scales. We investigate the degradation in the accuracy in determining some cosmological parameters that arises when assuming commonly employed assumptions and approximations.
	To do this, we investigate the level of accuracy provided by different models for the galaxy two point correlation function for measurements of the growth of structure and Primordial non-Gaussianity for two possible Stage IV-like surveys. While high accuracy is obviously always to be preferred, approximations are usually taken in order to simplify otherwise complicated and computationally expensive calculations using detailed modelings. However, this is valid only up to the point where the accuracy is not tampering with the scientific output of high precision measurements.
	
	After quantifying how error bars and best fit values are affected by different standard approximations, we propose a new hybrid model that retains enough accuracy being still much simpler than the full model.
	
	We obtain this by correctly accounting for the geometry of the system formed by the observer and the galaxy pairs, and in particular keeping the information given by the radial part of the correlation function, which has been often neglected (but see recent works on it~\cite{Raccanelli:2023fle, Raccanelli:2023zkj, Gao:2023rmo}).

	\section{Wide angle 3D galaxy clustering including PRIMORDIAL NON-GAUSSIANITY}
	\label{sec:wa}
\begin{figure}
		\centering
		\includegraphics[width=0.4\textwidth]{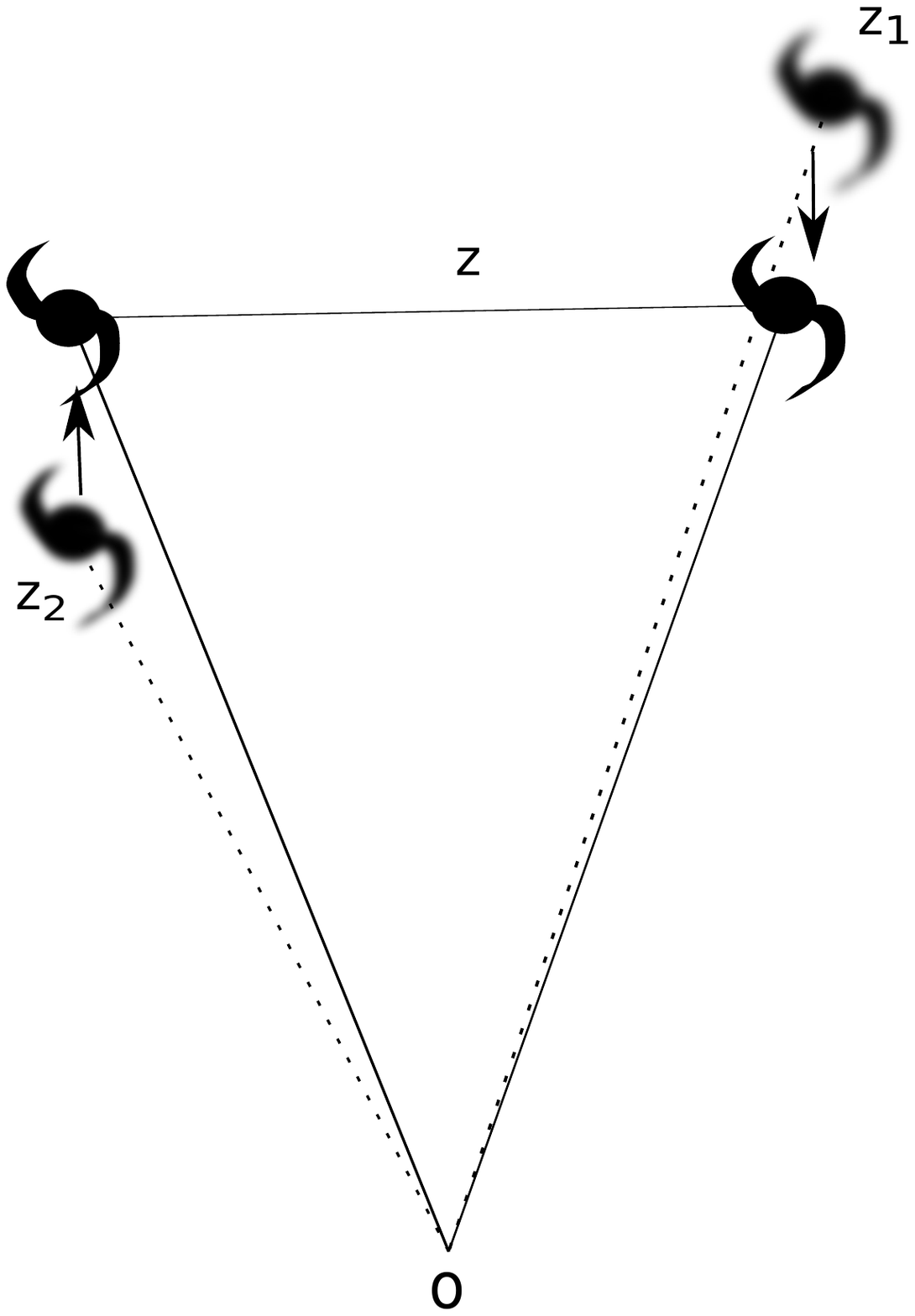}
		\includegraphics[width=0.4\textwidth]{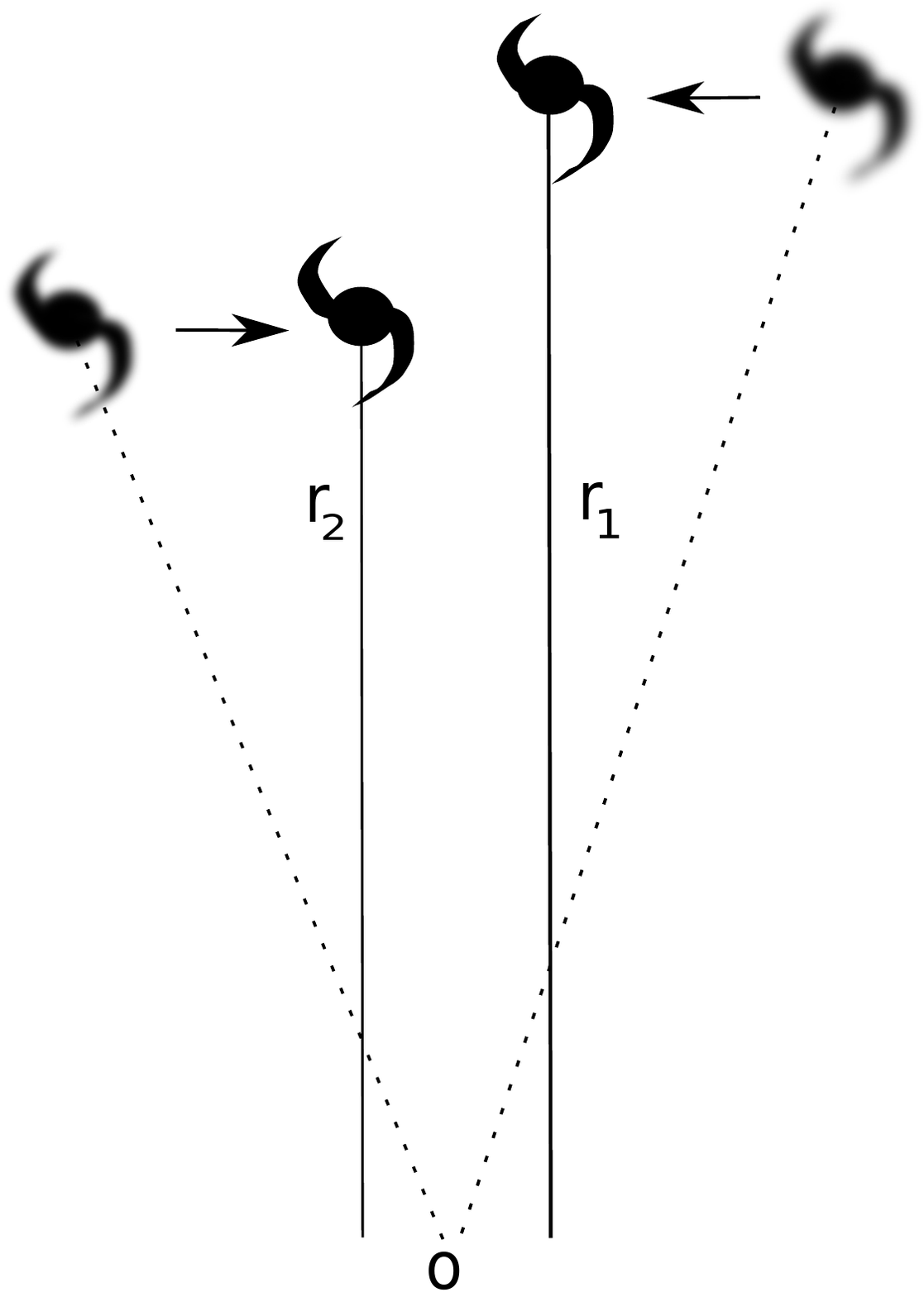}
		\caption{({\it Left}) Flat sky approximation and ({\it Right}) plane parallel approximation}
		\label{fig:approxs}
	\end{figure}

\begin{figure}
		\includegraphics[width=.4\textwidth]{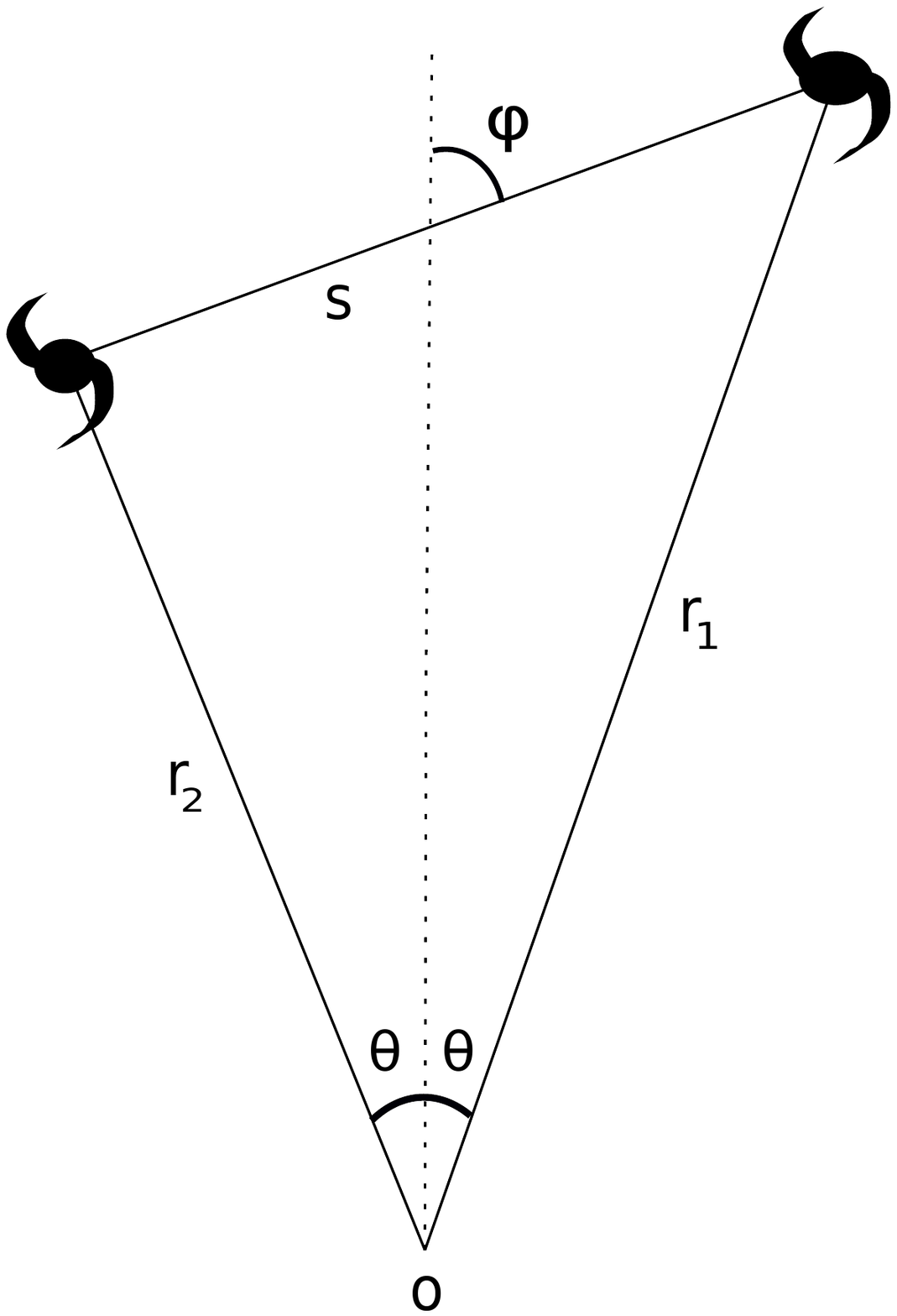}
		\caption{Coordinates system adopted in this paper}
		\label{fig:coordinates}
	\end{figure}

	When modeling the galaxy two point correlation function (2PCF), one can safely use the plane parallel or flat sky approximations in the cases where the area covered by a galaxy survey is very small compared to the distance between the observer and the sources.
	Note that the two approximations are similar but not strictly the same. The plane parallel (or distant observer) approximation, showed in the left panel of Fig.~\ref{fig:approxs}, assumes that since the linear separation between two galaxies is much smaller than their distance from us, we can treat the two line of sights connecting them to the observer as parallel to each other, and we can approximate to $0$ the opening angle between them  (for the geometry of the problem treated here, see Figure~\ref{fig:coordinates}). The flat sky limit, in the right panel of Fig.~\ref{fig:approxs}, implies that we are flattening the sky in the radial direction (i.e.,~we see all the galaxies at the same redshift), failing to account for the full 3-dimensionality of the problem.
	The flat sky approximation can be extended to include radial separations as recently investigated in~\cite{Raccanelli:2023fle,Raccanelli:2023zkj,Gao:2023rmo} in the context of angular and Fourier space correlations. We will see later how this can be implemented in our configuration space treatment in order to improve upon the flat sky modeling, without necessarily implement the fully exact calculation.
	In what follows, we refer to the flat sky model as a formalism that employs both the plane-parallel and the flat sky approximations.
	
	Given that the forthcoming generation of galaxy surveys will probe larger cosmological volumes, both in area and in redshift range,
	it is worth investigating if these approximations fail to model with enough accuracy the statistical properties of galaxy clustering. In fact, for future wide surveys such as SPHEREx~\cite{Dore:2014cca}, Euclid~\cite{Euclid:2024yrr,EUCLID:2011zbd,Euclid:2019clj}, DESI~\cite{DESI:2016fyo,DESI:2024mwx} and the Vera Rubin Observatory~\cite{LSSTDarkEnergyScience:2012kar}, that will observe galaxy pairs with very large angular separations, it will be crucial to check the accuracy of the modeling employed.
	Scales probed by such surveys have not been accessible until now, so, in order to correctly extract all the cosmological information from the data, it will be necessary to pair newly achieved precision with matching accuracy.
	
	Similarly, very deep surveys like the proposed ATLAS Explorer~\cite{Wang2018}, SIRMOS~\cite{XXX} and MegaMapper~\cite{Schlegel:2022vrv}, require to account for the full 3D geometry of the system, including galaxy pairs with large radial separation; moreover, high-$z$ galaxies have larger redshift uncertainties compared to low-$z$ ones, and this can result in having very thick bins in the radial direction, making the flat sky approximation less accurate.
	For this reason, there have been several studies trying to understand the importance of correct radial and transverse modeling for future surveys.
	
	Since we do not observe true galaxy positions, but we infer distances from measured redshifts, the coherent comoving galaxy velocities due to the growth of structure leads to measurable anisotropic clustering. The radial component of the peculiar velocity of each galaxy will be misinterpreted by the observer as cosmological in origin and so one defines the space in which we observe the position of galaxies as the redshift space, and the space in which galaxies would have been if their position was determined only by the Hubble flow as the real space (this is the standard redshift space distortions theory, for details see e.g.,~\cite{Kaiser1987, Hamilton1997,Fisher:1993pz,Cole:1993kh}). At linear order, the relation between the observed galaxy overdensity and the real space one can be written, in Fourier space, as:
	
	\begin{equation}
		\label{eq:overd}
		\delta_{obs,g}(\mathbf{k},z)=D(z)\big[b(z)+\mu^2 f(z) -i\alpha(z)\mu f(z)k^{-1}\big]\delta_m(\mathbf{k}) \; ,
	\end{equation}
	where $\mu =\cos{\phi}$, with $\phi$ the angle with the line of sight and $f$ is the growth rate parameter.
	In this work, as we want to evaluate best fit shifts on some example parameters, we parameterize the growth of structure as~\cite{Linder2005}:
	\begin{equation}
		f(z)=\Omega_{m}(z)^{\gamma} \; ,
	\end{equation}
	which has been found to be a good fit for many models; $\gamma$ is called the growth index and in the standard $\Lambda$CDM+GR scenario $\gamma\sim0.55$.
	
	In this work, since we are considering only large scales ($\gtrsim 30$ Mpc/$h$ see Sect.~\ref{sec:fisher} for details), we will work at linear level, assuming a linear bias relation between the galaxies and matter overdensities~\cite{Kaiser:1984sw,Bardeen:1985tr,Mo:1995cs,Desjacques:2016bnm}:
	
	\begin{equation}\label{eq:bias}
		\delta_g(\mathbf{k},z)\sim b(z)\delta_m(\mathbf{k})\; .
	\end{equation}
	
	While the first two terms in Equation~\eqref{eq:overd} are the usual Kaiser term~\cite{Kaiser1987, Hamilton1997}, the contribution proportional to $\alpha(z)$ is related to the fact that local overdensities around each galaxy can affect the apparent movement of galaxies when going from real- to redshift-space, and can be thought of as a Doppler velocity term~\cite{Szalay:1997cc, Papai:2008bd, Raccanelli:2010hk, Raccanelli:2016avd, Abramo:2017xnp,Andrianomena:2018aad} .
	In the Newtonian limit we have~\cite{Raccanelli:2016avd}: 
	\begin{equation}
		\alpha(z)=\frac{2}{\chi}+H(z)\frac{d\ln N(z)}{dz}\;,
	\end{equation}\label{eq:doppler_term}
	where $N(z)$ is the galaxy redshift distribution and $\chi(z)$ is the comoving distance of the source.
	A more general expression for the $\alpha$-term is needed if one wants to include local general relativistic effects that can be non negligible at very large scales~\cite{Jeong:2011as,Bertacca:2012tp,Raccanelli:2016avd,Semenzato:2024rlc}, but including such corrections is beyond the scope of this paper.

	
	\subsection{Primordial non-Gaussianity}
	\label{sec:PnG}
	The standard single-field models of slow-roll inflation predict a small deviation from Gaussian primordial curvature perturbations, while non-standard scenarios such as multi field models allow for a larger level of non-Gaussianity. Such deviations from Gaussianity can be parameterized by the dimensionless parameter $\fnl$ as~\cite{Salopek1990, Komatsu2001, Verde:2000vr}:
	\begin{equation}
		\label{eq:fnl}
		\Phi_{\rm NG}=\phi+\fnl\left[\phi^2-\langle\phi^2\rangle\right] \; ,
	\end{equation}
	where $\Phi$ denotes Bardeen's gauge-invariant potential, which, on sub-horizon scales, reduces to the usual Newtonian gravitational potential.
	Here $\phi$ is the Gaussian random field and the second term accounts for deviations from Gaussianity.
	The primordial bispectrum generated by Primordial non-Gaussianity (PNG) couples small scale density perturbations with large scale potential modes, affecting dark matter halo clustering at late time so that a non-zero $\fnl$ in Equation~\eqref{eq:fnl} introduces a scale-dependent modification of the halo bias~\cite{Matarrese2008,Matarrese:2000iz,Verde:2000vr}; we can write the total non-Gaussian bias in the case of local shape PNG as~\cite{Matarrese2008,Dalal2007}:
	\begin{equation}
		\label{eq:ng-bias}
		b_{\rm NG}(z, k) = b_G(z) + b_{\phi} \fnl \frac{3 \Omega_{0m}H_0^2}{2k^2T(k)g(z)} \; ,
	\end{equation}
	
	where $b_G(z)$ is the usual scale-independent bias of Equation~\eqref{eq:bias}, calculated assuming Gaussian initial conditions, $T(k)$ is the matter transfer function and $g(z)$ is the growth factor normalized to the scale factor during the matter-dominated era.
	Measurements of $f_{\rm NL}$ depend on the knowledge of the local PNG bias parameter $b_{\phi}$, that in turns depend on the assembly bias, potentially representing a problem for reaching high-precision measurements of $f_{\rm NL}$ (see e.g.,~\cite{Barreira:2020ekm,Barreira:2022sey}). However,~\cite{Fondi:2023egm} extended the work of~\cite{reid2010non}, showing how assembly bias can be accurately modeled and priors on $b_{\phi}$ can be imposed by considering the conditional halo mass function. Since this topic is currently the subject of ongoing discussions, and for the sake of simplicity, here we assume universality relation and a fair sample of halos. 
	
	Equation~\eqref{eq:overd}, including a non-Gaussian bias, becomes:
	\begin{equation}\label{eq:overd_ng}
		\delta_{obs,g}(\mathbf{k},z)=D(z)\bigg(b(z)+\mu^2 f(z) -i\mu f(z)\alpha(z)k^{-1}+\Delta b(z)k^{-2}\frac{1}{T(k)}\bigg)\delta_m(\mathbf{k}) \; ,
	\end{equation}
	where:
	\begin{equation}
		\Delta b(z)=[b(z)-1]\fnl\delta_{\rm ec} \frac{3 \Omega_{0m}H_0^2}{g(z)}\; , 
	\end{equation}
	and $\delta_{\rm ec}$ is the critical value of the matter overdensity for elliptical collapse in an Einstein-de Sitter universe taken to be $\delta_{\rm ec}=1.686$~\cite{Sheth:1999su}. In general, a modification $\delta_c\rightarrow q\delta_c$ is often assumed, in which $q$ has the role of a fudge factor that accounts for example for the fact that the actual definition of halos depends on the halo finder
	algorithm considered, which in turn affects the halo number counts and thus their bias. For simplicity in the following we assume that $q=1$.
	In this work we focus only on the so-called local type non-Gaussianity, as customary for analyses of 2-point correlations, as constraints on shapes other than the local type are considerably weakened (see e.g.,~\cite{Giannantonio2011}); for a review of different shapes of PNG, see e.g.,~\cite{Babich:2004gb}.
	
	\subsubsection{Flat sky approximation}
	The simplest model for the 2PCF employs both the flat sky and the plane parallel approximations, and galaxies are treated as they are all at the same redshift. In this scenario the 2PCF will depend only on the modulus of the linear separation between the two sources $s$, the orientation angle $\phi$  between $\mathbf{s}$ and the radial direction $\mathbf{\hat{n}}$ and the effective redshift at which the 2PCF is calculated. In this work, we neglect Doppler terms in what we call the "flat sky" model in order the study the impact of neglecting such terms in the modelling (see Appendix~\ref{app:doppler} for details). In this scenario we have the standard result~\cite{Hamilton1997}:
	\begin{equation}\label{eq:fsgauss}
		\begin{gathered}
			\xi_{pp}(s,\phi,z)=D(z)^2\bigg[\left(b(z)^2+\frac{2b(z)f(z)}{3}+\frac{f(z)^2}{5}\right)\xi^0_0(s)+\\
			\left(-\frac{4}{3}b(z)f(z)-\frac{4}{7}f(z)^2\right)\xi^0_2(s)\mathcal{L}_2(\cos(\phi))+\frac{8}{35}f(z)^2\xi^0_2(s)\mathcal{L}_4(\cos(\phi))\bigg]\; ,
		\end{gathered}
	\end{equation}
	where $\mathcal{L_n}$ are the the Legendre polynomials of order $n$,
	with the added non-Gaussian contributions:
	\begin{equation}\label{eq:fsnongauss}
		\xi_{pp,ng}(s,\phi,z)=D(z)^2\left[\left(\Delta b(z)^2\xi^4_{0,ng}(s)+2\Delta b(z)\bigg(b(z)+\frac{f(z)}{3}\right)\xi^2_{0.ng}(s)-\frac{4}{3}\Delta b(z)f(z)\xi^2_{2,ng}(s)\mathcal{L}(\cos(\phi))\right]\; .
	\end{equation}
	
	Note that in the flat sky approximation, the RSD kernels of the observed 2PCF in the $i$-th redshift bin are often computed at a fixed $z_{\mathrm{eff},i}$. Here for simplicity we assume $z_{\mathrm{eff},i}=(z_{i,max}+z_{i,min})/2$, while in real data analyses the effective redshift is calculated applying specific weights to each galaxy (see e.g.,~\cite{Bautista:2020ahg,Sanchez:2013uxa,Samushia:2011cs,eBOSS:2019gbd,eBOSS:2020qek}). 
	
	\subsection{Wide angle 3D correlation}
	\label{sec:tripolar}
	In order to drop the flat sky and plane parallel approximation, one can develop a formalism that allows to retain the full 3-dimensionality of the physics involved.
	
	Following the approach proposed in~\cite{Szalay:1997cc} and developed in~\cite{Szapudi:2004gh, Papai:2008bd, Raccanelli:2010hk}, we define the spherical transforms of the matter overdensity based on Legendre polynomials $\mathcal{L}_{\ell}$:
	\begin{equation}
		\mathcal{A}_{\ell}^{n}(\mathbf{x})=\int \frac{d^{3} k}{(2 \pi)^{3}}(i k)^{-n} \mathcal{L}_{\ell}(\mu) \exp (i \mathbf{k} \cdot \mathbf{x}) \delta(\mathbf{k})\; .
	\end{equation}
	For the non Gaussian term we have:
	\begin{equation}
		\Tilde{\mathcal{A}}_{0}^{2}(\mathbf{x})=\int \frac{d^{3} k}{(2 \pi)^{3}}(i k)^{-2} \frac{1}{T(k)}\mathcal{L}_{0}(\mu) \exp (i \mathbf{k} \cdot \mathbf{x}) \delta(\mathbf{k})\; .
	\end{equation}
	We can therefore write, in configuration space:
	\begin{equation}\label{newterms}
		\begin{aligned}
			\quad\delta_{g,obs}(\mathbf{x},z) =D(z)\left[\left(b(z)+\frac{1}{3} f(z)\right)\mathcal{A}_{0}^{0}(\mathbf{x})
			+\frac{2}{3} f(z) \mathcal{A}_{2}^{0}(\mathbf{x})+\alpha(z) f(z) \mathcal{A}_{1}^{1}(\mathbf{x})-\Delta b(z) \Tilde{\mathcal{A}}_{0}^{2}(\mathbf{x})\right]\; ,
		\end{aligned}
	\end{equation}
	so that the 2-point correlator of the spherically transformed matter overdensity is:
	\begin{equation}
		\begin{aligned}
			\left\langle\mathcal{A}_{\ell_{1}}^{n_{1}}\left(\mathbf{x}_{1},z_1\right) \mathcal{A}_{\ell_{2}}^{n_{2}}(\mathbf{x}_{2},z_2\right)\rangle=D(z_1)D(z_2)\times
			(-1)^{\ell_{2}} \int \frac{d^{3} k}{(2 \pi)^{3}}(i k)^{-\left(n_{1}+n_{2}\right)} \mathcal{L}_{\ell_{1}}(\hat{\mathbf{k}} \cdot \mathbf{n}_{1}) \mathcal{L}_{\ell_{2}}(\hat{\mathbf{k}} \cdot \mathbf{n}_{2}) 
			e^{i \mathbf{k} \cdot \mathbf{s} }P(k)\; ,
		\end{aligned}
	\end{equation}
	where $s$ is the linear comoving separation between the two galaxies and $P(k)$ is the matter power spectrum. 
	
	Expanding $\mathcal{L}_{\ell}$ and $\exp (i \mathbf{k} \cdot \mathbf{s})$ in spherical harmonics and applying the Gaunt integral, we obtain:
	\begin{equation}
		\left\langle\mathcal{A}_{\ell_{1}}^{n_{1}}\left(\mathbf{x}_{1},z_1\right) \mathcal{A}_{\ell_{2}}^{n_{2}}\left(\mathbf{x}_{2},z_2\right)\right\rangle=D(z_1)D(z_2)\times
		\sum_{L}(-1)^{\ell_{2}} i^{L-n_{1}-n_{2}}(2L+1)\left(\begin{array}{ccc}
			\ell_{1} & \ell_{2} & L \\
			0 & 0 & 0
		\end{array}\right)
		\times S_{\ell_{1} \ell_{2} L}\left(\mathbf{n}_{1}, \mathbf{n}_{2}, \mathbf{n}_{12}\right) \xi_{L}^{n_{1}+n_{2}}\left(s\right)\; ,
	\end{equation}
	for $\left|\ell_{1}-\ell_{2}\right| \leq L \leq \ell_{1}+\ell_{2}$, that is an expansion of the correlator in the tripolar basis functions~\cite{Szalay:1997cc}:
	\begin{equation}
		S_{\ell_{1} \ell_{2} L}\left(\mathbf{n}_{1}, \mathbf{n}_{2}, \mathbf{n}_{12}\right)=\left[\frac{(4 \pi)^{3}}{\left(2 \ell_{1}+1\right)\left(2 \ell_{2}+1\right)(2 L+1)}\right]^{1 / 2}
		\times \sum_{m_{1}, m_{2}, M}\left(\begin{array}{ccc}
			\ell_{1} & \ell_{2} & L \\
			m_{1} & m_{2} & M
		\end{array}\right) Y_{\ell_{1} m_{1}}\left(\mathbf{n}_{1}\right) Y_{\ell_{2} m_{2}}\left(\mathbf{n}_{2}\right) Y_{L M}\left(\mathbf{n}_{12}\right) \; ,
	\end{equation}
	for $\ell_{1} \leq m_{1} \leq \ell_{1},\ell_{2} \leq m_{2} \leq \ell_{2}$ and $L \leq M \leq L$ and with coefficients:
	\begin{equation}
		\label{coef}
		\xi_{L}^{n}\left(s \right)=\int \frac{d k}{2 \pi^{2}} k^{2-n} j_{L}(s k) P\left(k\right) \; .
	\end{equation}
	In the non-Gaussian case, the additional contribution is:
	\begin{equation}
		\label{eq:coef_ng}
		\xi_{L,ng}^{n}\left(s \right)=\int \frac{d k}{2 \pi^{2}} \frac{k^{2-n}}{T(k)^m} j_{L}(s k) P\left(k\right) \; ,
	\end{equation}
	where $m=2$ for $n=4$ and $m=1$ in the other cases.
	
	The $k-$behaviour of each correlator is governed by the index $n$.
	When $n=4$, we require an IR cutoff $k_{min}=10^{-4}\sim\frac{H_{0}}{2}$ as the integrand of Equation~\eqref{eq:ng_coeff} becomes power-law divergent. This will not impact our results since pushing the lower extreme of the integration below $k_{min}$ only adds an overall additive normalization to $\xi$ which is unobservable~\cite{Bertacca:2012tp}. Moreover~\cite{Desjacques:2020zue} showed that the sensitivity of the local PNG term on $k_{min}$ is well below cosmic variance. 
	Note that the correlators $\langle\mathcal{A}_1^1,\mathcal{A}_1^1\rangle$,$\langle\mathcal{\tilde{A}}_0^2,\mathcal{A}_0^0\rangle$, $\langle\mathcal{A}_0^0,\mathcal{\tilde{A}}_0^2\rangle$ have the same $\propto k^{-2}$ behaviour. This means that in principle the contributions coming from the Doppler terms can mimic an effective non-Gaussianity~\cite{Raccanelli:2016avd,Abramo:2017xnp} in the galaxy 2-point correlators.
	
	The tripolar decomposition of the observed two point galaxy correlation function is then:
	\begin{align}
		\label{eq:expansion}
		\xi(s,\theta,\phi,z_1,z_2)
		&=D(z_1)D(z_2) \sum_{\ell_{1}, \ell_{2}, L, n} B_n^{{\ell_{1} \ell_{2} L}}(z_1,z_2) S_{\ell_{1} \ell_{2} L}(\theta,\phi) \xi_{L}^{n}(s) \; .
	\end{align}
	
	The complete set of coefficients $B_n^{l_1,l_2,L}(z_1,z_2)$ for the Gaussian case can be found in literature, and for completeness we report them here in Appendix~\ref{app:coeffs}. For the non-Gaussian contributions, we have:
	\begin{equation}\label{eq:ng_coeff}
		\begin{gathered}
			\xi_{ng}(s,\theta,\phi,z_1,z_2)=D(z_1)D(z_2)\bigg\{\Delta b(z_1)\Delta b(z_2)\xi^4_{0,ng}(s)+\left[\Delta b(z_1)\left(b(z_2)+\frac{f(z_2)}{3}\right)+\Delta b(z_2)\left(b(z_1)+\frac{f(z_1)}{3}\right)\right]\xi^2_{0.ng}(s)\\
			-\frac{2\sqrt{5}}{3}\left(S_{022}(\theta,\phi)\Delta b(z_1)f(z_2)+S_{202}(\theta,\phi)\Delta b(z_2)f(z_1)\right)\xi^2_{2,ng}(s)\\
			+\xi^3_{1,ng}(s)\sqrt{3}\left(-S_{011}(\theta,\phi)\Delta b(z_1)         \alpha(z_2)f(z_2)+S_{101}(\theta,\phi)\Delta b(z_2)
			\alpha(z_1)f(z_1)\right)\bigg\}\; .
		\end{gathered}
	\end{equation}
	
	We see that the coefficients $\xi_m^l$ depend only on the linear separation between the two sources $s$ and on the cosmological parameters; the full 3D geometry of the system is encoded in the tripolar spherical harmonics basis $S_{\ell_{1} \ell_{2} L}(\theta,\phi)$, while the terms that describe deviations from Gaussianity, the growth of structure, and all the projection effects are included in the coefficients $B_{{\ell_{1} \ell_{2} L}}(z_1,z_2)$.

	\section{Observed multipoles  of the galaxy 2PCF in redshift space}
	\label{sec:multipoles}
	Since the three unit vectors that define the position of the two galaxies w.r.t. the observer are on the same plane, the redshift space correlation function depends only on the shape and size of the triangle formed by the observer and the two sources. This triangle is uniquely determined by specifying any three of its variables, such as three sides, three angles, two angles and one side, etc.
	In this work we express our calculations using the linear separation between the two galaxies, \{$s,\theta$\} that is the half of the opening angle, and $\phi$, the angle between the bisector of the opening angle and $s$. The $\hat{z}$ axis is aligned with the bisector of $\theta$ that defines the $\phi=0$ direction. The system of coordinates is shown in Fig.~\ref{fig:coordinates}.

	In order to fully describe the observable with only three degrees of freedom $(s,\phi,\theta)$, we have to remove the redshift dependence ($z_1,z_2$) in Equation~\eqref{eq:expansion}.
	We can write the redshift dependence of the 2PCF inverting the following relations~\cite{Papai:2008bd}:
	\begin{equation}
		\label{eq:coord}
		r_1=\frac{\sin(\phi+\theta)}{\sin(2\theta)}s, \quad
		r_2=\frac{\sin(\phi-\theta)}{\sin(2\theta)}s \; ,
	\end{equation}
	in order to express $z_1(r_1(s,\theta,\phi))$ and $z_2(r_2(s,\theta,\phi))$.
	At a given linear separation $s$ and orientation angle $\phi$, galaxy pairs might be observed with different opening angles $\theta$. For instance, pairs that are at a certain distance from the observer are seen under a smaller $\theta$ w.r.t. pairs that are closer. The resulting observed 2PCF is then averaged over all pairs with fixed $s$ and $\phi$, as in~\cite{Yoo:2013zga}: 
	\begin{equation}
		\label{eq:xitheta}
		\xi_{\mathrm{obs}}(s,\phi)=\frac{\int^{\theta_{max}}_ {\theta_{min}}d\theta N(s,\theta,\phi)\xi(s,\theta,\phi)}{\int^{\theta_{max}}_ {\theta_{min}}d\theta N(s,\theta,\phi)}\; ,
	\end{equation}
	where $N(s,\theta,\phi)$ is a weight function that accounts for the distribution of pairs (see Sect.~\ref{sec:ntheta} for more details). 
	
	Since we perform observations on the spherical sky, galaxy clustering analyses are often carried out by expanding the 2PCF in Legendre polynomials~\cite{Hamilton1997} as each multipole contains different physical information about the underlying clustering, so it is convenient to focus on them for our modeling.
	We can expand the 2PCF with Legendre polynomials as~\cite{Hamilton1997}:
	\begin{equation}\label{eq:Xi}
		\xi_{L}(s)=(2L+1)\int_0^{\pi/2}\xi_{\mathrm{obs}}(s,\phi)\sin(\phi)\mathcal{L}_{L}(\cos(\phi))d\phi\; ,
	\end{equation}
	where we used the fact that for single tracer analyses the 2PCF is symmetric w.r.t the transformation $\phi\rightarrow \pi-\phi$.


	\subsection{Radial distribution}
	\label{sec:ntheta}
	The integral in Equation~\eqref{eq:xitheta} over the opening angle $\theta$ allows to account for the full 3D structure of the problem. 
	However, we have to consider the full distributions of galaxies in \{$s,\theta,\phi$\}, which in general can be not uniform (see e.g.,~\cite{Samushia:2011cs}).
	This needs to be considered by applying a weight function,
	$\xi_{\mathrm{obs}}(s,\theta,\phi)=N(s,\theta,\phi)\xi(s,\theta,\phi)$.
	In a real data analysis, those distributions will be extracted from the observed catalog. In a theoretical forecast as the one we present here, the weightings depend on the probability to find a galaxy in a particular position, which is connected to the specific redshift distribution of the survey, and we need to account for the survey geometry.
	The finiteness of the bin width translates into a limited range of $\theta$ under which the different pairs can be observed in that bin, for each linear separation $s$ and orientation $\phi$.
	In order to predict $N(s,\theta,\phi)$, we simulate mock catalogs of galaxies by populating the survey volume with particles distributed according to the specific $N(z)$ of that survey and tabulate the probability distribution as function of the coordinates \{$s$,$\theta$,$\phi$\}, seen by the observer. Of course this is an idealized scenario, because in practical surveys the observed region of the sky is discontinuous, the mask is non trivial and the angular selection function is not uniform.
	
	In this work we are interested in theoretical forecasts, so in order to get physical insights we limit ourselves to the idealized scenario where the probability distributions of a triangle configuration is uniquely specified, given $N(z)$ and $f_{\rm sky}$. 
	As an example mock survey we use a SPHEREx-like survey ($f_{sky}=0.75$), with two different binning schemes:
	
	-\textit{Thick bins} scenario:  $0<z<4.6$, bins with $0<z<1$ have $\Delta z=0.2$, while bins with $1<z<4.6$ have $\Delta z=0.6$ (this is the standard binning scheme for the SPHEREx survey~\cite{Dore:2014cca}) 
	
	-\textit{Thin bins} scenario: $0<z<2$, where all bins have equal width $\Delta z=0.2$.
	
\begin{figure}
		\includegraphics[width=0.47\textwidth]{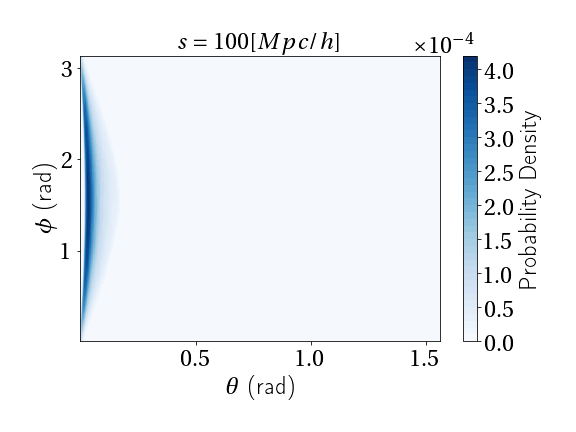}
		\includegraphics[width=0.47\textwidth]{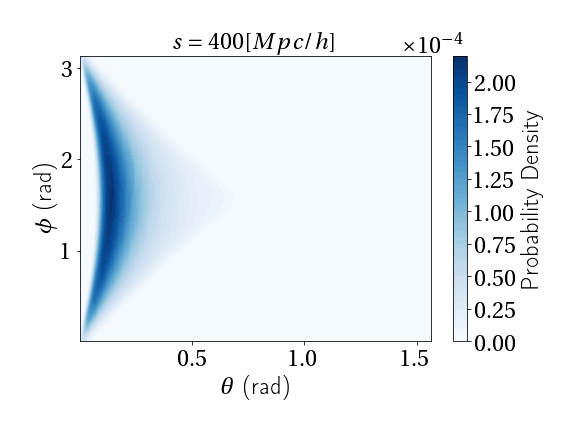}
		\caption{$N(s,\theta,\phi)$ distributions for fixed values of linear separations $s$. The distribution is normalized as $\int N(s,\theta,\phi)d\theta d\phi=1$ for each value of $s$.} 
		\label{fig:n_theta_phi}
	\end{figure}

	We use the SPHEREx galaxy sample with $\sigma(z)/(1+z)\leq0.03-0.1$, in order to minimize the shot noise contribution; we do not account for the error introduced by the redshift uncertainty, and we checked that they do not propagate differently to different modelings. 
	In Fig.~\ref{fig:n_theta_phi} we show the probability distribution $N(s,\theta,\phi)$ obtained with the procedure outlined above for two fixed values of the linear separations. We see that the majority of the galaxies are seen under a small opening angle even for large separations; this fact will be important to understand the results of Section~\ref{sec:utc}.
	
	If we are expanding the 2PCF in Legendre polynomials, they must be an orthonormal basis independent from the chosen coordinate system; this condition is broken in the 3D wide angle case, as the triangle formed by the observer and the two sources closes (and so the galaxy pair is actually observable) only when $\phi$ is larger than $\theta$ 
	(see also~\cite{Raccanelli:2013dza}). In order to deal with this issue in the full 3D modeling, one can either follow~\cite{Raccanelli:2013dza} and modify the expansion, or limit the maximum linear separation in order to allow the integral on $\phi$ to be performed over the entire $\mu=\{-1, 1\}$ range.
	Following the latter, this limitation requires imposing an upper limit on the maximum linear separation $s_{\mathrm{max},i}$ observed in the $i$-th redshift bin.
	We checked that, for every bin, correlations with separation larger than $s_{\mathrm{max}}$ are noise dominated and so they will not add any further significant information to the analysis. Therefore, we limit the linear separations we consider to values that satisfy this condition.
	
	Moreover, in the wide angle 3D model we can drop the approximation that the galaxies are located at the same (effective) redshift $z_{\mathrm{eff}}$. In fact, using two angles,~i.e., the orientation $\phi$ and the opening angle $\theta$ (as in Fig.~\ref{fig:coordinates}) allows us to account properly for the 3D geometry of the system and to model correctly the position of each galaxy inside the redshift bin and the distributions of the sources with respect to the observer.  
	For this reason, in the flat sky scenario, the more pairs we are ``mis-measuring'' (wrongly identifying their redshift with $z_\mathrm{eff}$), the worse we are modeling the correlation function.


	\section{Fisher Analysis}
	\label{sec:fisher}
	\subsection{Fisher matrix and best fit shifts}
	\label{sec:fm}
	We perform a Fisher forecast analysis~\cite{ Vogeley:1996xu,Tegmark:1997rp} to investigate how wide angle 3D corrections affect measurements of the growth index and Primordial non-Gaussianity, for some example future galaxy surveys.
	We then study how incorrectly modeling the 2PCF with the (standard) flat sky approximation will cause a mis-estimation of the inferred best fit values of the growth index and $f_{\rm NL}$.
	In what follows, we assume perfect knowledge of the cosmological parameters at their fiducial values obtained by the Planck Collaboration~\cite{Planck2018}, as we are interested not in the exact forecasting of cosmological parameters, but on the impact of using the flat sky approximation.
	We, instead, leave the bias parameters in each bin free to vary.
	The Fisher matrix element is, for each redshift bin:
	\begin{equation}\label{eq:fisher}
		F_{a,b}=\sum_{i, j} \frac{\partial\tilde{\xi}(s_i)}{\partial \theta_a} \operatorname{Cov}^{-1}\left(s_i, s_j\right) \frac{\partial\tilde{\xi}(s_j)}{\partial \theta_b}\; ,
	\end{equation}
	where $\tilde{\xi}(s)=\{\xi_{0}(s_i),\xi_{0}(s_{i+1})...\xi_{0}(s_{N}),\xi_{2}(s_i),\xi_{2}(s_{i+1})...\xi_{2}(s_{N})\}$ is the vector containing the theoretical monopole and quadrupole of the 2PCF.
	
	Since we use only linear scales in the analysis, we adopt a conservative approach similar to the one proposed by~\cite{Giannantonio2011} to determine the minimum linear separation at each redshift.
	We fix the maximum scale $k=0.15 h$/Mpc at $z=0$ and we keep the variance of the matter perturbations constant at each redshift to find $k_{NL}(z)$ 
	imposing the condition that at $z=0$, $\sigma^2(z)=0.36$.
	We fixed and absolute maximum value of $k_{NL}=0.3$, and checked that a less conservative approach does not significantly change the results of this work.
	
	Using the flat sky approximation in the theoretical model will lead to a wrong estimation for the parameters that best fit the data. We can model the shift in this best fit value of the parameter $\Theta$  as~\cite{Bernal:2020pwq,Raccanelli:2017kht}:

	\begin{equation}\label{eq:shift}
		\Delta \Theta=\left(\sum_m F_m\right)^{-1}\left[\sum_{m, i, j}\nabla_\Theta\tilde{\xi}^{pp}(s_i)_{m }\left(\operatorname{Cov}^{pp}(s_i,s_j)^{-1}\right)_m\left(\tilde{\xi}^{wa}(s_j)_{m }-\tilde{\xi}^{pp}(s_j)_{m }\right)\right]\; ,
	\end{equation}
	
	where the sum is performed over the $m$ redshift bins; this is valid since we assume that measurements in different non-overlapping redshift bins are not correlated; this is valid in the case where integrated projection effects are neglected (see e.g.,~\cite{Raccanelli:2013gja}).
	
	\subsection{Theoretical covariance}
	\label{sec:covariance}
	We estimate the theoretical covariance for the multipoles of the correlation function for each redshift bin following the procedure outlined in~\cite{Bonvin:2015kuc,Tansella:2018sld}: 
	
	\begin{equation}
		\begin{aligned}
			\operatorname{Cov}_{\ell, \ell',i}(s, s') = \frac{i^{\ell - \ell'}}{V_i} \Bigg[& \frac{(2 \ell + 1)^2}{2 \pi \bar{n_i}^2 s^2} \delta(s - s') \delta_{\ell \ell'} 
			+ D(z_i)^2\sum_L \left(\begin{array}{ccc}
				\ell & \ell' & L \\
				0 & 0 & 0
			\end{array}\right)^2 \times \\
			& \times \left(\frac{1}{\bar{n_i}}\int c_L(k,z_i) \mathcal{G}_{\ell \ell'}(k,s, s')dk + D(z_i)^2\int \tilde{c}_L(k,z_i)\mathcal{D}_{\ell \ell'}(k,s, s')dk\right)\Bigg] \; ,
		\end{aligned}
		\label{eq:covar}
	\end{equation}
	
	where $\Bar{n}_i$ is the average galaxy number density in the $i$-th redshift bin centered at redshift $z_i$ of volume $V_i$ and:
	\begin{equation}
		\begin{aligned}
			& \mathcal{G}_{\ell \ell^{\prime}}\left(s, s^{\prime}\right)=\frac{2(2 \ell+1)\left(2 \ell^{\prime}+1\right)}{\pi^2} k^2 P(k) j_{\ell}(k s) j_{\ell^{\prime}}\left(k s^{\prime}\right) \, , \\
			& \mathcal{D}_{\ell \ell^{\prime}}\left(s, s^{\prime}\right)=\frac{(2 \ell+1)\left(2 \ell^{\prime}+1\right)}{\pi^2} k^2 P^2(k) j_{\ell}(k s) j_{\ell^{\prime}}\left(k s^{\prime}\right)\; ,
		\end{aligned}
	\end{equation}
	and the coefficients $c_L$ (where $L=0,2,4$) are obtained by expanding the galaxy power spectrum in redshift space in Legendre polynomials $P(k,\mu)=c_0P(k)+c_2P(k)\mathcal{L}_2(\mu)+c_4P(k)\mathcal{L}_4(\mu)$, with~\cite{Hamilton1997}:
	\begin{equation}
		\begin{aligned}
			c_0 & =b^2+\frac{2}{3} b f+\frac{f^2}{5} \, , \\
			c_2 & =\frac{4}{3} b f+\frac{4}{7} f^2 \, , \\
			c_4 & =\frac{8}{35} f^2 \; ,
		\end{aligned}
	\end{equation}
	where the redshift and $k$ dependencies are omitted for brevity.
	The coefficients $\tilde{c}_L$ obtained for the expansion of $(P(k,\mu)/P(k))^2$ in Legendre polynomials (i.e.,~($c_0+c_2\mathcal{L}_2(\mu)+c_4\mathcal{L}_4(\mu))^2=\sum_L\tilde{c}_L\mathcal{L}(\mu)$ are:
	\begin{equation}
		\begin{aligned}
			& \tilde{c}_0=c_0^2+\frac{c_2^2}{5}+\frac{c_4^2}{9}, \\
			& \tilde{c}_2=\frac{2}{7} c_2\left(7 c_0+c_2\right)+\frac{4}{7} c_2 c_4+\frac{100}{693} c_4^2, \\
			& \tilde{c}_4=\frac{18}{35} c_2^2+2 c_0 c_4+\frac{40}{77} c_2 c_4+\frac{162}{1001} c_4^2\; .
		\end{aligned}
	\end{equation}

	As we can see from Equation~\eqref{eq:covar}, theoretical errors are calculated assuming deviations from Gaussianity only at the power spectrum level (i.e.,~in the scale dependent bias) neglecting contributions coming from the connected 4-point correlation function.
	
	Note that since the expression for the theoretical covariance is given in the flat sky approximation the constraints obtained for the wide angle model are only an estimation as for a proper exact calculations one need to substitute in Equation~\ref{eq:fisher} the wide angle expression of the covariance. On the other hand the results obtained for the best fit shift are exact as the covariance in Equation~\eqref{eq:shift} is the one calculated assuming a wrong modelling~i.e., the flat sky approximation.

	\section{Modeling}
	\label{sec:modeling}
	Following all the definitions and considerations above, we now define the different models we will test for the analyses of galaxy clustering. After having introduced the two extremes of full 3D wide angle and the Kaiser, flat-sky plane-parallel ones, we here define some intermediate models that promise better accuracy but with considerably simpler expressions, and test how accurately they can retrieve some cosmological parameters.
	For simplicity, in the following we will refer to the ``flat sky'' model for the case where we use the flat sky, the plane parallel and the effective redshift (i.e.,~all the sources inside a redshift bin are evaluated at the same redshift) approximations.
	
	\begin{figure}
		\includegraphics[width=.49\textwidth]{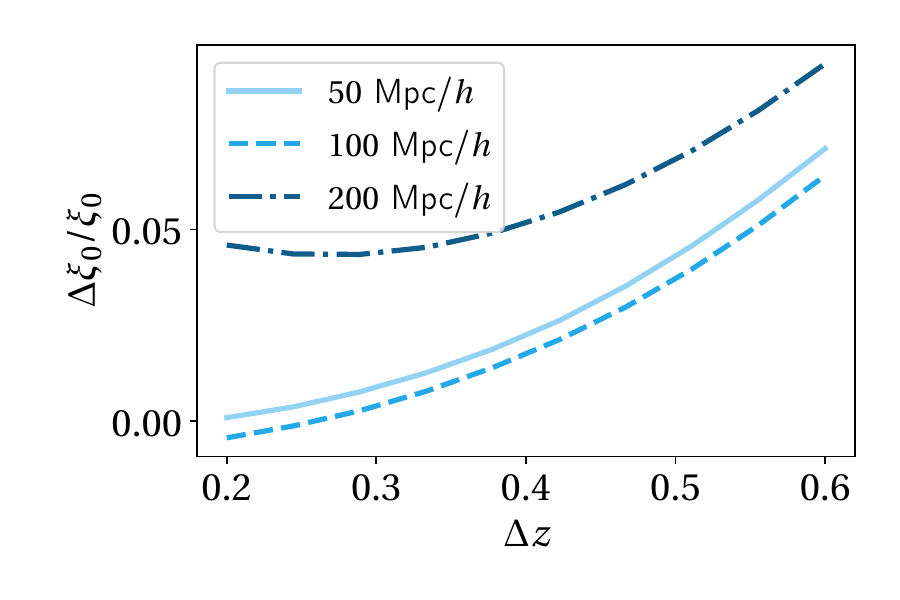}
		\includegraphics[width=.49\textwidth]{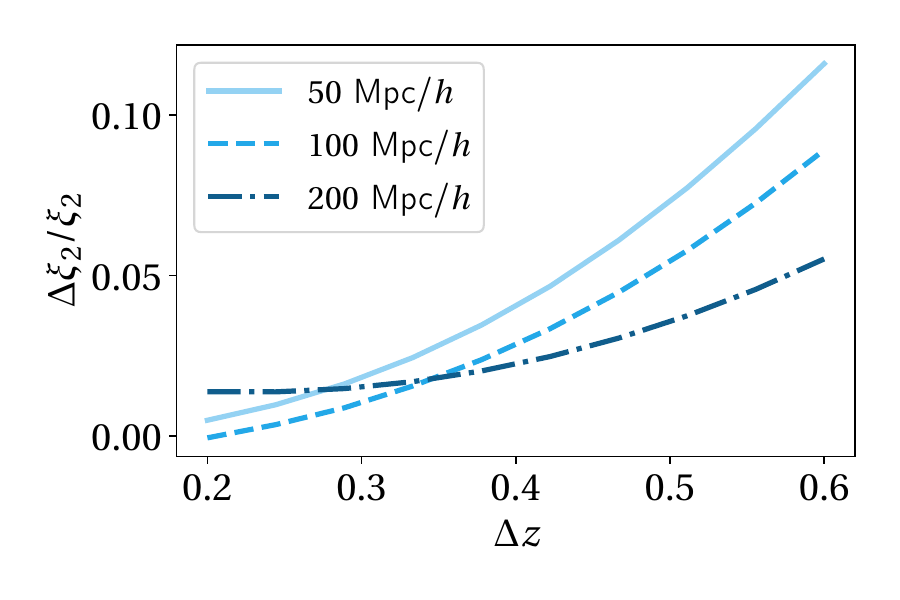}
		\caption{Difference between the wide angle 3D model and the flat sky approximation as a function of the width if the redshift bin for three linear separation for the monopole (left) and the quadrupole (right).}
		\label{fig:diff}
	\end{figure}
	
	In Fig.~\ref{fig:diff} we plot the difference between the wide angle 3D model and the flat sky one as a function of the width of the redshift bin, for three fixed linear separations; we see that the difference between the two models is larger for thick bins, where flattening all the galaxies to an average redshift is less accurate, as expected. 
	
	One can try to improve the modeling of the 2PCF by dropping some of the approximations explained above, still without using the full wide angle 3D formalism, in order to minimize the theoretical systematics that will lead to biased results in the final measurements.
	
	\subsection{Radial averaging}
	\label{sec:zavg}
	One can argue that fixing the radial position of the sources inside a redshift bin to some $z_{\mathrm{eff}}$ will lead us to a loss of information on the radial clustering of galaxies, especially if the bin is very thick. Such a compression of the full 3D information is in principle unjustified as quantities such as the bias, the growth factor evolve within the bin range, and the clustering along the radial directions can be non negligible. In fact, results from previous sections show that this can be one of the main sources of inaccuracy.
	
	To overcome this issue, one can average the flat sky model over each redshift bin and weighting for the redshift distribution of the sources. The observed correlation function in this case reads:
	\begin{equation}
		\label{eq:avg}
		\xi_{\mathrm{obs}}(s,\phi)=\frac{\int_{z_{\mathrm{min}}}^{z_{\mathrm{max}}}N(z)\xi_{pp}(s,\phi,z)dz}{\int_{z_{\mathrm{min}}}^{z_{\mathrm{max}}}N(z)dz}\; .
	\end{equation}
	
	However, note that this procedure allows us to account for all the equal time correlations (i.e.,~correlations of sources at the same redshift) in a specific redshift bin, but the flatness of the sky, by definition, prevents us to model correctly the unequal time correlations that are present in the wide angle 3D model. Such correlations have been recently modelled by~\cite{Raccanelli:2023fle,Raccanelli:2023zkj}, where it is showed that there can be non negligible corrections to the equal time galaxy power spectrum for radial correlations.
	
\begin{figure}
		\includegraphics[width=.48\textwidth]{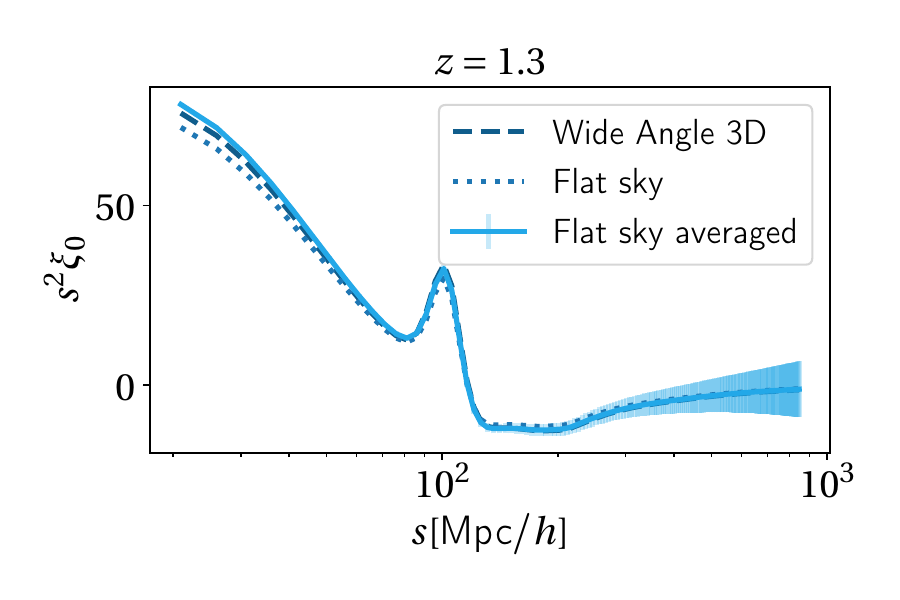}
		\includegraphics[width=.48\textwidth]{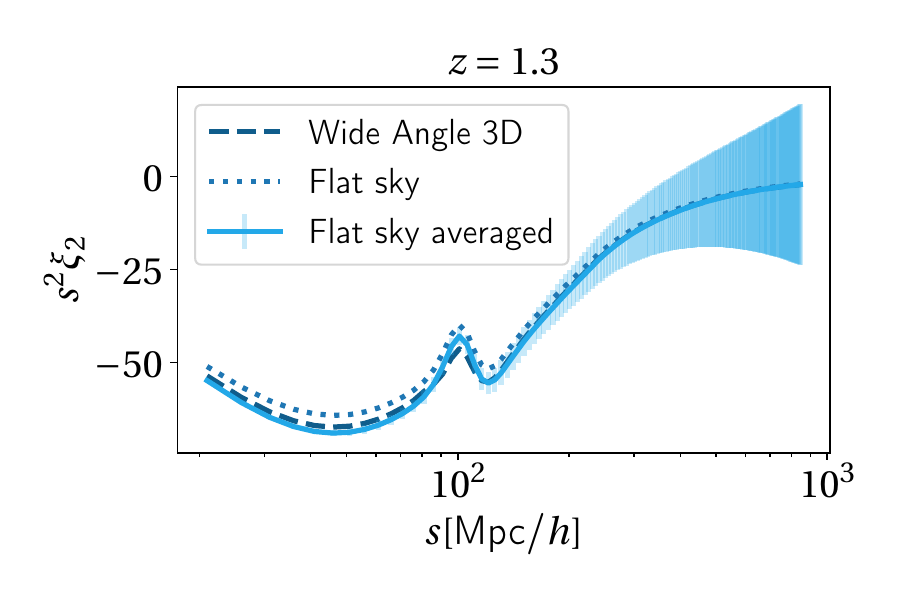}
		\includegraphics[width=.48\textwidth]{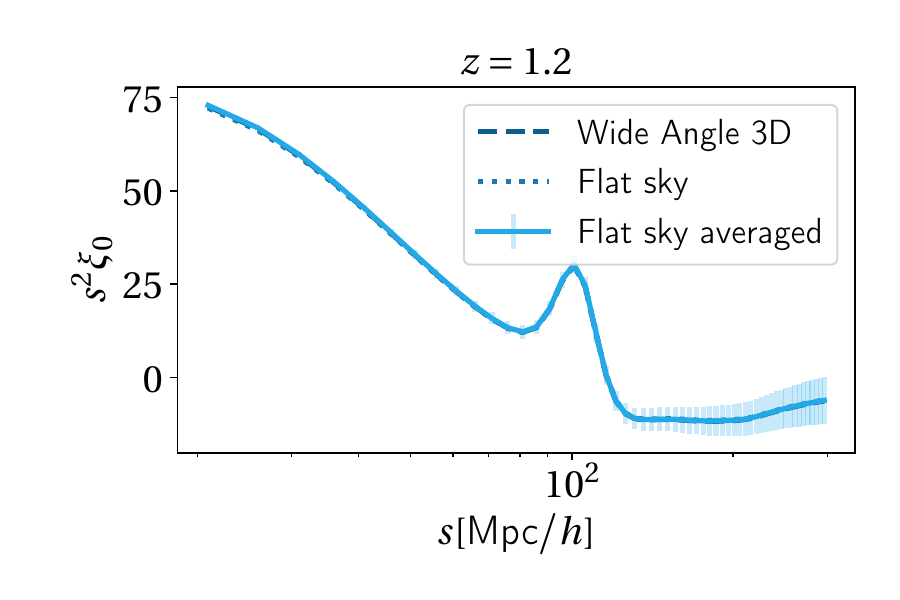}
		\includegraphics[width=.48\textwidth]{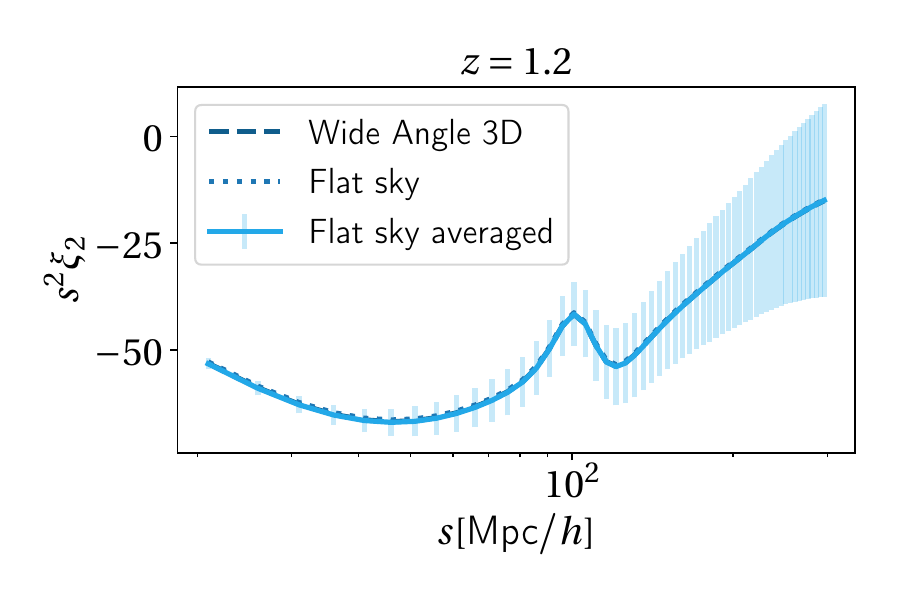}
		\caption{Monopole and quadrupole of the galaxy 2PCF in redshift space for different models in the thick bins scenario (upper panels) and in the thin bin scenario (lower panels). Continuous lines: flat sky averaged model. Dashed lines: wide angle 3D model. Dotted lines: flat sky model}
		\label{fig:xiplots}
	\end{figure}

	In Fig.~\ref{fig:xiplots} we look again at the comparison between the full 3D wide angle case and the flat sky one, but this time for the full shape of the correlation function, and including the radial averaging case.
	We can notice already an improvement with respect to the flat sky case, especially for thick bins, again confirming expectations.
	The Figure shows the monopole and quadrupole of the galaxy 2PCF in redshift space for the full wide angle 3D model (dashed lines), the flat sky averaged one (Equation~\eqref{eq:avg}) (continuous line) and the flat sky modeling (dotted lines) for the mock surveys used in this work: thick bin scenario (upper panels), thin bin scenario (lower panels).
	
	\subsection{Radial 3D model}
	\label{sec:utc}
	The next model we investigate, in the context of a simplified version of the full 3D wide angle that retains as much accuracy as possible, is what we call the ``radial 3D'' model.
	
	As shown in Fig.~\ref{fig:n_theta_phi}, in practice galaxies are too far away w.r.t. the observer to be seen under large aperture angle $\theta$, so one can study what happens if we take the limit $\theta=0$ in the tripolar basis coefficients $S_{\ell_1,\ell_2,L}$ in Equation~\eqref{eq:expansion}. At the same time, however, we keep the dependence on $\theta$ in the redshift dependent quantities (i.e.,~in the coefficients $B_{n}^{\ell_{1},\ell_{2},\ell_{L}}$) in order to maintain the 3-dimensionality of the problem.
	In this scenario the 2PCF in redshift space can be expressed with a simpler formula:
	\begin{equation}
		\label{eq:radial}
		\xi(s,\theta,\phi)=D_1D_2\sum_n F_{L,n}(s,\theta,\phi)\xi_{L}^n(s)\mathcal{L}_L(\cos\phi) \; .
	\end{equation}
	The explicit expression is given by:
	\begin{equation}
		\begin{gathered}
			\xi(s,\theta,\phi)=D_1D_2\bigg\{\left(\mathcal{F}_1\mathcal{F}_2+\frac{4}{45}f_1 f_2 \right)\xi^0_0-\left[\frac{2}{3}\bigg(\mathcal{F}_1f_2+\mathcal{F}_2f_1\bigg)+\frac{8}{63}f_1f_2\right]\mathcal{L}_2(\cos\phi)\xi^0_2+\frac{8}{35}f_1f_2 \mathcal{L}_4(\cos\phi)\xi^0_4+\\
			\left[\alpha_1f_1\mathcal{F}_2-\alpha_2f_2\mathcal{F}_1+\frac{4}{15}f_1f_2\left(\alpha_1-\alpha_2\right)\right]\mathcal{L}_1(\cos\phi)\xi^1_1+\left[\frac{2}{5}f_1f_2(\alpha_2-\alpha_1)\right]\mathcal{L}_3(\cos\phi)\xi^1_3+\frac{\alpha_1\alpha_2f_1f_2}{3}\left(\xi^2_0-2\mathcal{L}_2(\cos\phi)\xi^2_2\right)\\
			\left(\Delta b_1\mathcal{F}_2+\Delta b_2\mathcal{F}_1 \right)\xi^2_{0,ng}+\Delta b_1\Delta b_2\xi^4_{0,ng}-\frac{2}{3}(\Delta b_1f_2+\Delta b_2f_1)\mathcal{L}_2(\cos\phi)\xi^2_{2,ng}+(\alpha_2f_2\Delta b_1-\alpha_1f_1\Delta b_2)\mathcal{L}_1(\cos\phi)\xi^3_{1,ng}\bigg\} \; ,
		\end{gathered}
	\end{equation}
	where $\mathcal{F}_i=b_i+1/3f_i$ and the subscript $i=1,2$ identifies different galaxies, 
	This expression, while obviously more complicated than the Kaiser formula, is still way simpler that then full 3D including wide angle terms. In Appendix~\ref{app:FF} we show the explicit expressions for the $F_{ij}$ terms.
	In this expression, all the quantities that depend on redshift ($\{f, b, \alpha\}$) will carry a dependence on the three variables $\{s, \theta, \phi\}$, and therefore require the use of the three degrees of freedom that are necessary to define the triangle formed by the observer and the galaxy pair.
	
	This simplified model is still able to capture the full 3-dimensionality of the problem, as the peculiar position of each sources inside the redshift bin is modelled correctly.

\begin{figure}
		\includegraphics[width=0.35\textwidth]{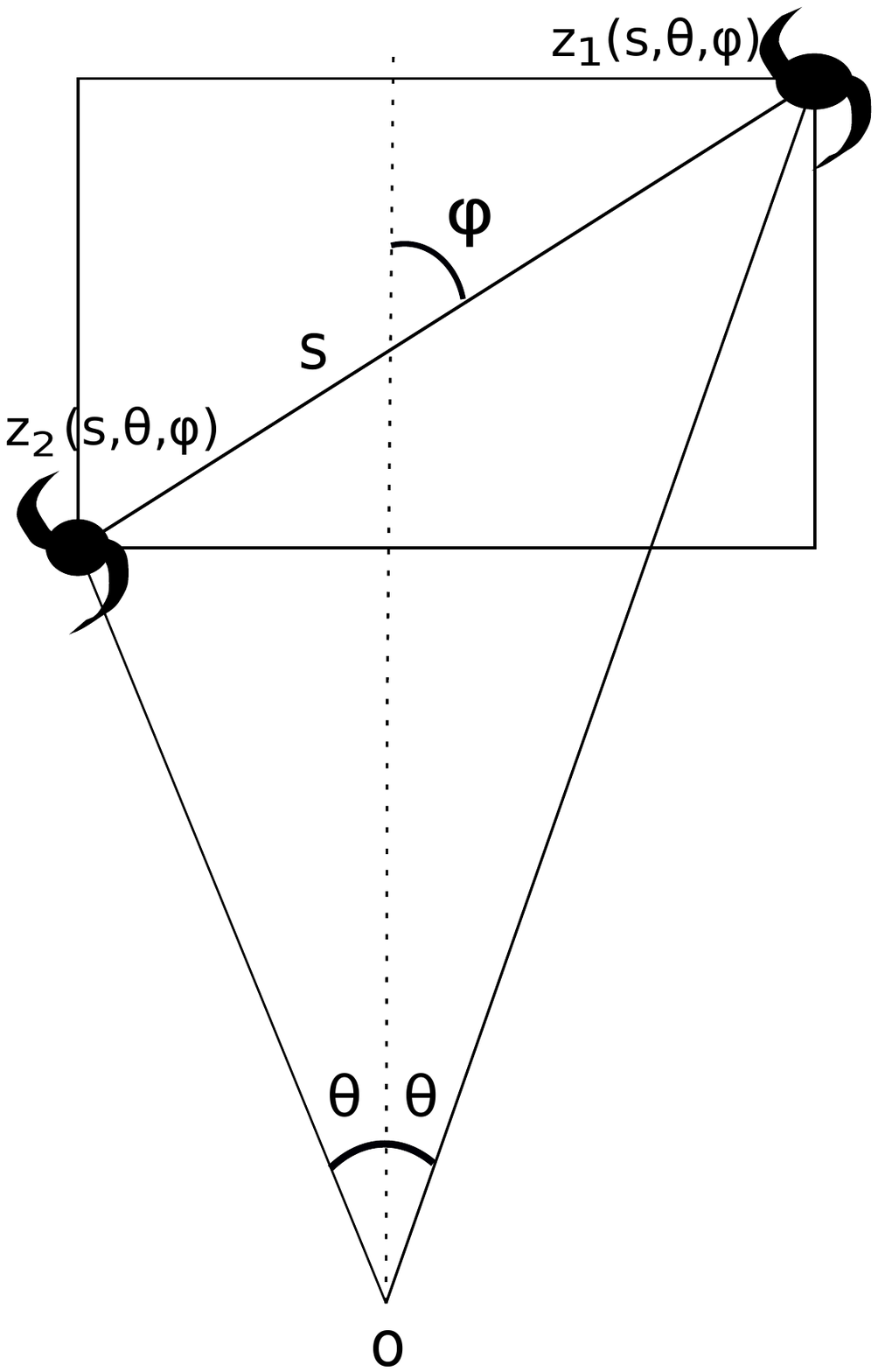}
		\caption{Illustration of the 3D radial setup.} 
		\label{fig:3Dradialbox}
	\end{figure}

	In fact, once we identify a box of cosmic volume defined by a given redshift range, since in the radial 3D model we preserve the three degrees of freedom of the full wide angle 3D formalism ($\{s,\theta,\phi\}$), we can correctly identify every triangle configuration formed by the observer and the sources in the box, while being able to weight every configuration for the probability of being observed, as described in details in Sect.~\ref{sec:ntheta}.
	In this way, we can properly account for unequal time correlators as the the full radial information is not lost as it is in the flat sky model; a visual illustration of the setup we use for this model is shown in Fig.~\ref{fig:3Dradialbox}.
	This allows the radial 3D model to be more accurate than the flat sky models while remaining simpler than the full wide angle 3D one also in terms of computational cost.
	Note that the radial 3D model is different w.r.t. the plane parallel (distant observer) one as in the latter case the aperture angle is not considered at all in the modeling. In the radial 3D model, instead, the line of sights are still treated as non parallel to each other and the $\theta$ coordinate is fundamental in order to correctly identify every triangle configuration, as explained above.
	
	\section{Results}
	\label{sec:results}
	In this work we present an estimation of the impact of taking some approximations used in galaxy clustering analyses instead of using a fully correct modeling.
	In particular, the results of our work quantify the impact of using the plane parallel and flat sky approximations in measurements of the growth of structure and Primordial non-Gaussianity.
	
	To interpret the results correctly, we must understand that the difference between the standard, so-called Kaiser flat sky model and the full, wide angle 3D one is twofold.
	In the wide angle 3D formalism there are terms proportional to the opening angle $\theta$ (which are not present in the standard modeling), which increase with $\theta$ and encode the dependence on the aperture angle between the correlated sources. These terms are present even if all the redshift dependent quantities in the 2PCF are evaluated at a fixed mean $z_\mathrm{eff}$ in each redshift bin (see~\cite{Szalay:1997cc}).
	
\begin{figure}
		\includegraphics[width=.48\columnwidth]{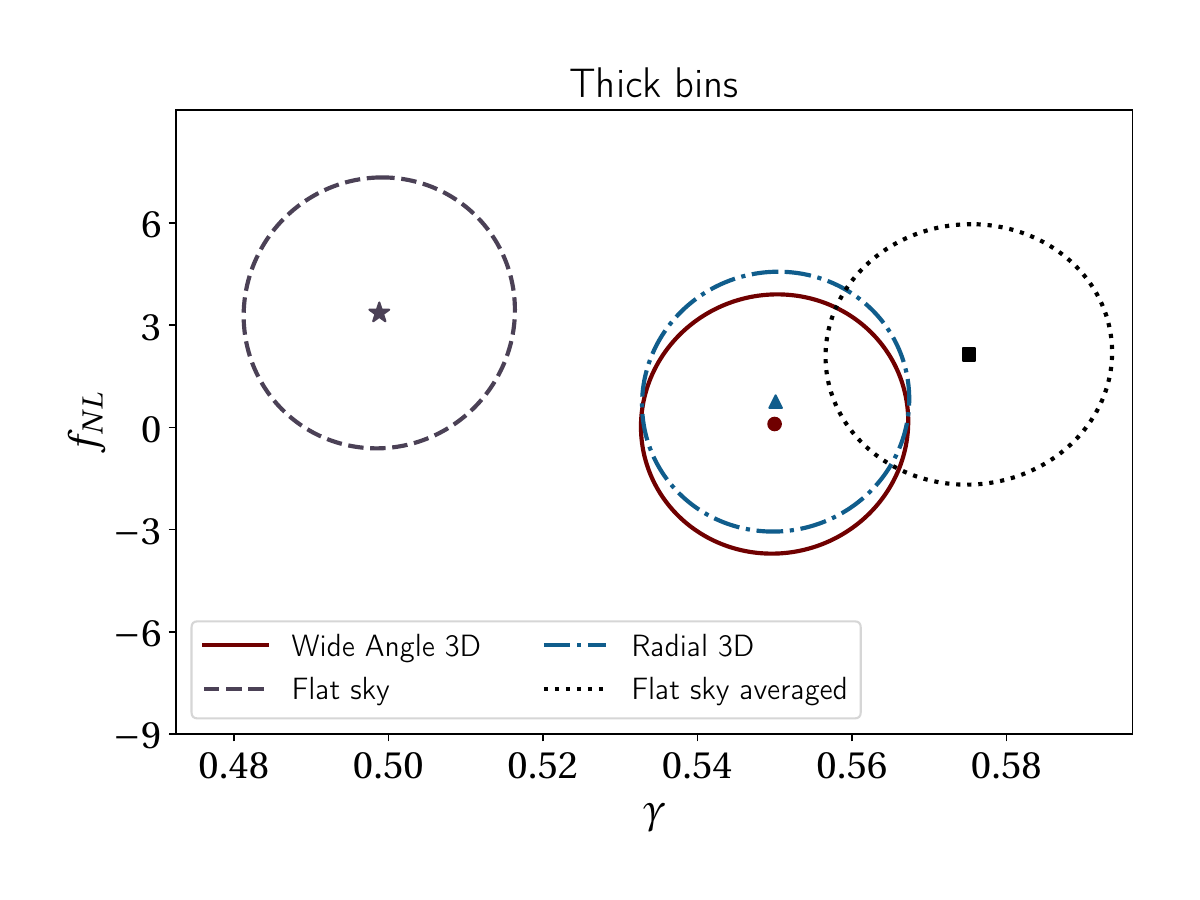}
		\includegraphics[width=.48\columnwidth]{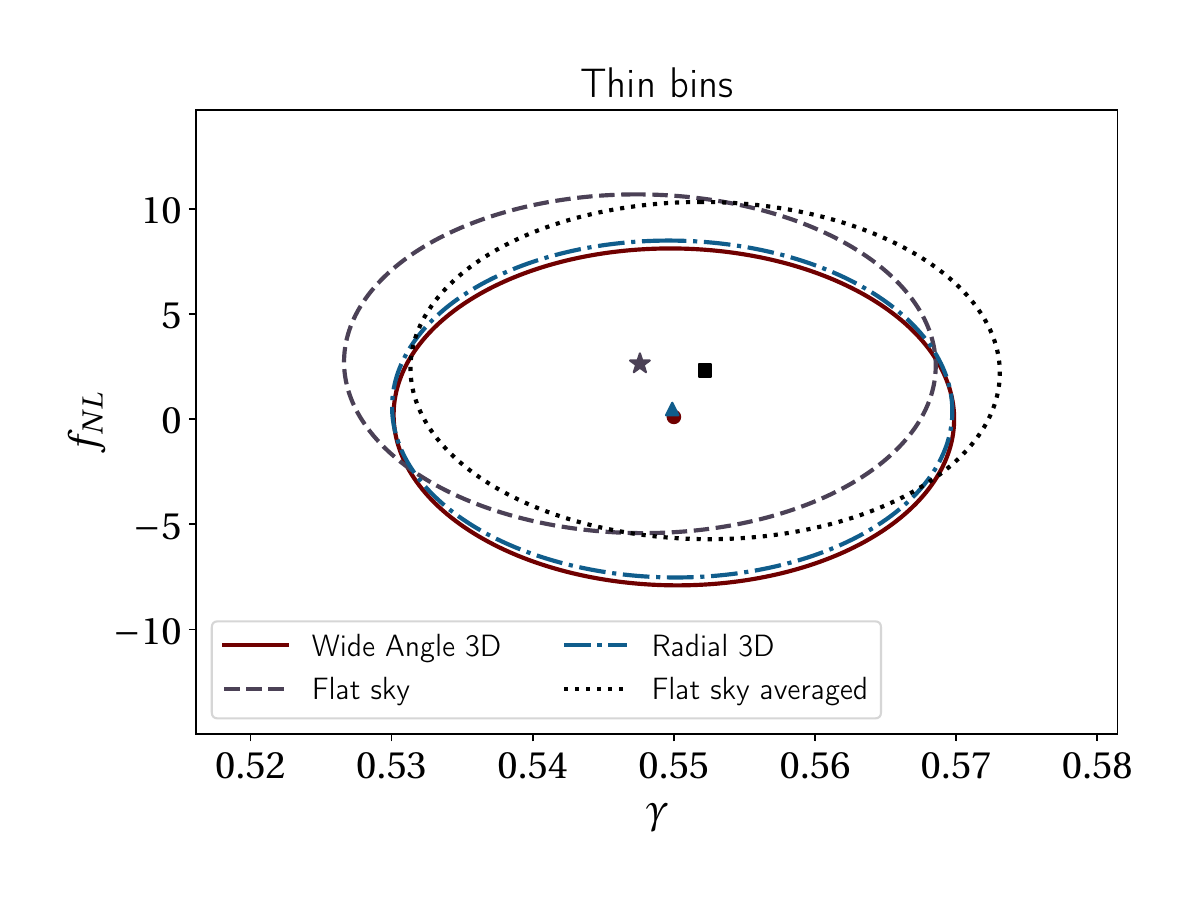}
		\caption{Best fits and $1\sigma$ limits of $\gamma$ and $\fnl$ in the case of the wide angle 3D model (red continuous line and circles), flat sky averaged (black dotted line and squares), flat sky (grey dashed line and stars) and the radial 3D model (blue dot-dashed line and triangles).
			Left panel is for the thick bins configuration, the right panel shows the thin bin scenario.
		}
		\label{fig:ellipses}
	\end{figure}

	In Fig.~\ref{fig:ellipses} we show the first of our main results, which is the predicted best fit and limits for different modelings.
	We show this for the $\gamma$ and $f_{\rm NL}$ parameters, comparing results when using the full 3D wide angle model or different approximations, namely the flat sky modeling (averaged and not) and a radial 3D model.
	In dashed grey and star is the forecast for the standard Kaiser modeling, using flat sky and one single effective redshift, while in solid red and circles we show the full 3D wide angle modeling. In dot-dashed blue and triangles we show the radial 3D model, and in dotted black and squares we show the flat sky averaged one. The left panel shows results for the SPHEREx-like thick bins scenario, while the right panel shows results for the thin bins configuration.
	
	We can see that, while constraints do not dramatically change for different models (the FoM changes by less than 10\%), the expected best fits change considerably for the thick bins.
	For the growth rate parameter, the expected high precision in future measurements and the fact that its redshift evolution is not properly captured when assuming all sources are at the same effective redshift, makes it that the best fit is considerably affected in the thick bins case.
	As expected, the effect becomes smaller when we consider thinner redshift bins.
	The flat sky model, even when averaged over $z$ within the bin, will get a $\sim 1\sigma$ shift in the best fit value of $\gamma$ in the case of the thick bin scenario.
	
	PNG, on the other hand, is less affected by the radial terms; however, its effect is larger at larger scales, and it is degenerate with geometry effects. For this reason, there is a shift also for the non-Gaussianity parameter.
	However, for the same reason, paired with the fact that the error bars are larger at large scales, the shift is approximately within one sigma for the thick bins case, and below that for the thin bins case.
	Note that such a shift, for future surveys, could be more relevant, especially when adding the bispectrum.
	\begin{figure}
		\centering
		\includegraphics[width=0.48\linewidth]{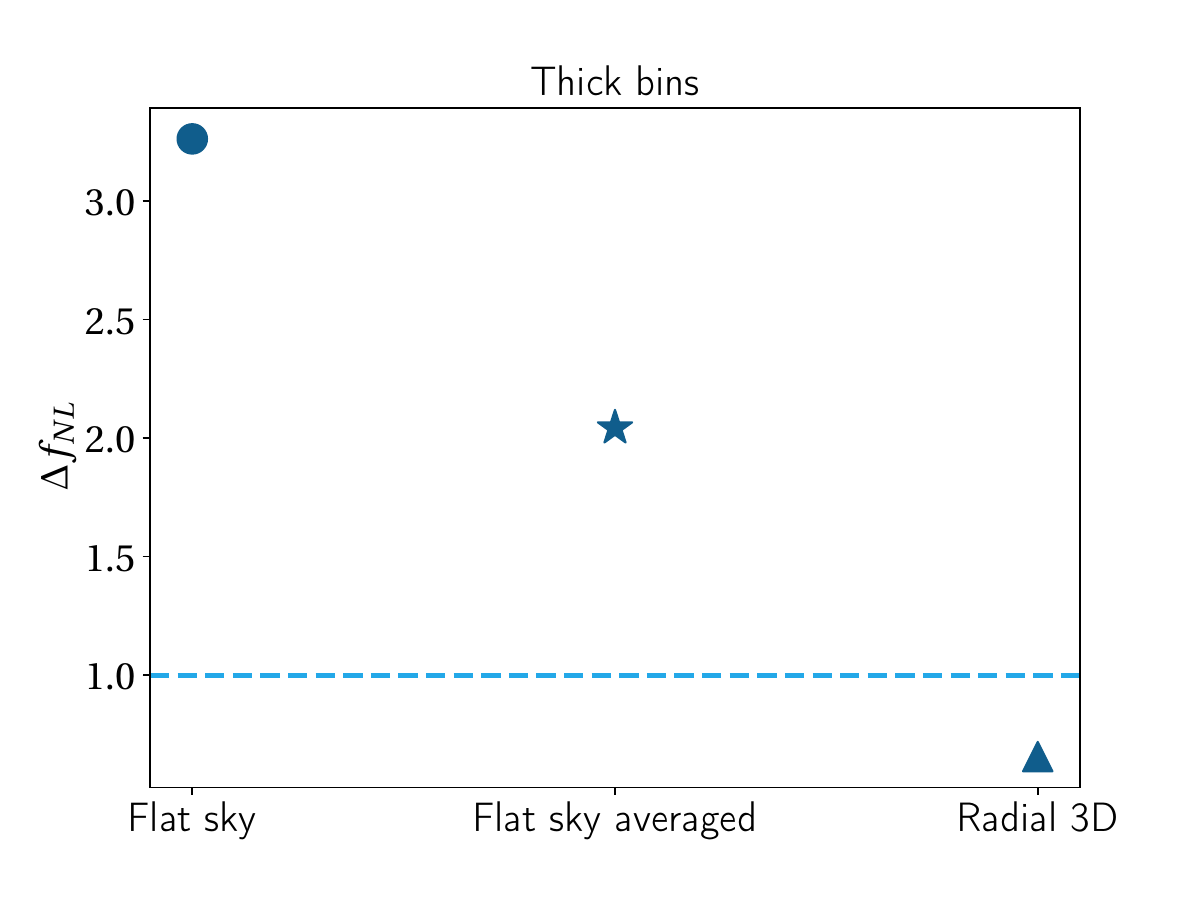}
		\includegraphics[width=0.48\linewidth]{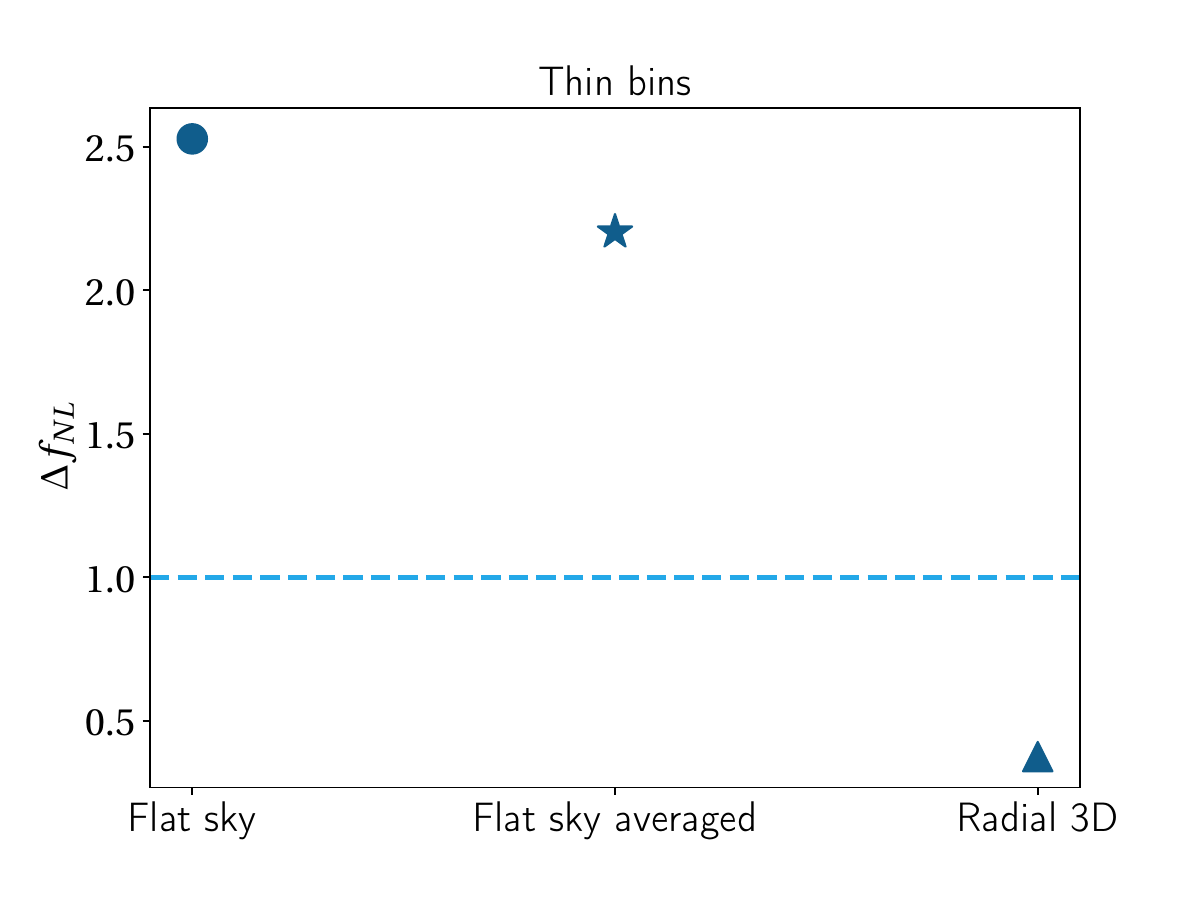}
		\caption{Shift in the best fit for $f_{\rm NL}$, for the different approximation models considered in this work. The blue dashed line indicates $f_{\rm NL}=1$, a target value for many future galaxy surveys.}
		\label{fig:fNLshift}
	\end{figure}
	
	In Figure~\ref{fig:fNLshift} we focus on the shift of the best fit for $f_{\rm NL}$. The plot shows the shift, for both cases and the various models considered here, induced by incorrect modeling. In dashed blue we show $f_{\rm NL}=1$, as this is the target value for many future surveys, given that it is generally the discriminant between single- and multi- field inflationary models~\cite{Alvarez:2014vva}. As we can see, while technically the shift on the inferred PNG parameter is below the $1-\sigma$-level, for theoretical reasons, and in anticipation of more precise measurements, it will still be very relevant for the flat sky models. The Radial 3D model, on the other hand, will only marginally bias the best fit.
	
	The fact that the Radial 3D model retains almost the same level of accuracy of the full wide angle 3D one is due to two main factors. The magnitude of the $\theta$-terms in the tripolar spherical harmonics coefficients increase at very large scales, where the error budget is large, and therefore the final impact on parameter estimation is reduced.
	As shown in Fig.~\ref{fig:n_theta_phi}, for a realistic survey scenario, galaxy pairs are far away from us so that there is little statistics for widely separated pairs.
	
	It is important to recognize, however, that this does not imply that it will be correct to use the plane parallel (or distant observer) approximation for modeling the observed position of each galaxy pair. The correct approach to describe the system comprising the observer and the two galaxies always requires three degrees of freedom (in this work they are taken to be $s,\theta,\phi$), rather than the two degrees offered by the plane parallel case, in order to account for the full 3-dimensionality of the problem.

	
	\subsection{nDGP}
	\label{sec:dgp}
	As a further application of our investigation, we look at how cosmological tests can be biased when using the flat sky approximation; in particular, we focus on tests of DGP~\cite{Dvali:2000hr}, an alternative cosmological model.
	In the original DGP model, the Universe is described by a 4D brane embedded in a 5D Minkowski spacetime. Matter is confined in the brane, while gravity propagates in the extra dimension above the cross-over scale $r_c$, which is the fundamental extra parameter of this model. General relativity is recovered by taking the cross-over scale to be much larger than the current Hubble scale, i.e.,~for $H_0 r_c \gg 1$. Constraints on this parameter are traditionally expressed in terms of the dimensionless quantity:
	\begin{equation}
		\Omega_{\mathrm{rc}} \equiv \frac{1}{4 r_c^2 H_0^2} \; .
	\end{equation}
	
	In the following we consider the "normal branch" of the DGP model (nDGP), where the background expansion history reproduces the $\Lambda \mathrm{CDM}$ one in a spatially-flat universe~\cite{Schmidt:2009sv} , since the so-called "self accelerating branch" which do not require the presence of a dark energy component to explain the expansion of the universe has been found to be unstable~\cite{Luty:2003vm,Koyama:2007za}. This is obtained by considering a dark energy component whose background dynamics exactly compensates the modified background evolution of the nDGP model so that deviations from $\Lambda$CDM take place only in the late-time evolution of perturbations.
	Constraints on the value of $\Omega_{\mathrm{rc}}$ has been recently investigated for example in~\cite{Raccanelli:2012gt, Liu:2021weo,Bosi:2023amu} and in~\cite{Piga:2022mge}, where they obtained an upper bound $\Omega_{\mathrm{rc}}\lesssim0.2$, analysing BOSS data.
	
	At linear order in perturbation theory, we can write the differential equation for the growth factor $D(a)$ in the nDGP model as~\cite{Schmidt:2009sv, Raccanelli:2012gt, Piga:2022mge}:
	\begin{equation}
		\frac{d^2 D(a)}{d \ln a^2}+\left(2+\frac{d \ln H}{d \ln a}\right) \frac{d D(a)}{d \ln a}-\frac{3}{2} \nu(a) \Omega_{\mathrm{m}, a}(a) D(a)=0\; .
		\label{eq:ndgp_evo}
	\end{equation}
	Assuming a $\Lambda$CDM background expansion rate, the difference with the $\Lambda \mathrm{CDM}$ case is captured by the function $\nu(a)$ in the last term, defined as:
	\begin{equation}
		\nu(a)=1+\frac{1}{3\beta(a)}\; , \quad \quad \quad 
		\beta(a) \equiv 1+\frac{H(a)}{H_0} \frac{1}{\sqrt{\Omega_{\mathrm{rc}}}}\left(1+\frac{a H^{\prime}(a)}{3 H(a)}\right) \; .
	\end{equation}
	which modifies the strength of the gravitational interaction. In the following we assume that $\nu \rightarrow 1$ (i.e.,~GR is recovered) at early times.
	
\begin{figure}
		\centering
		\includegraphics[width=.48\textwidth]{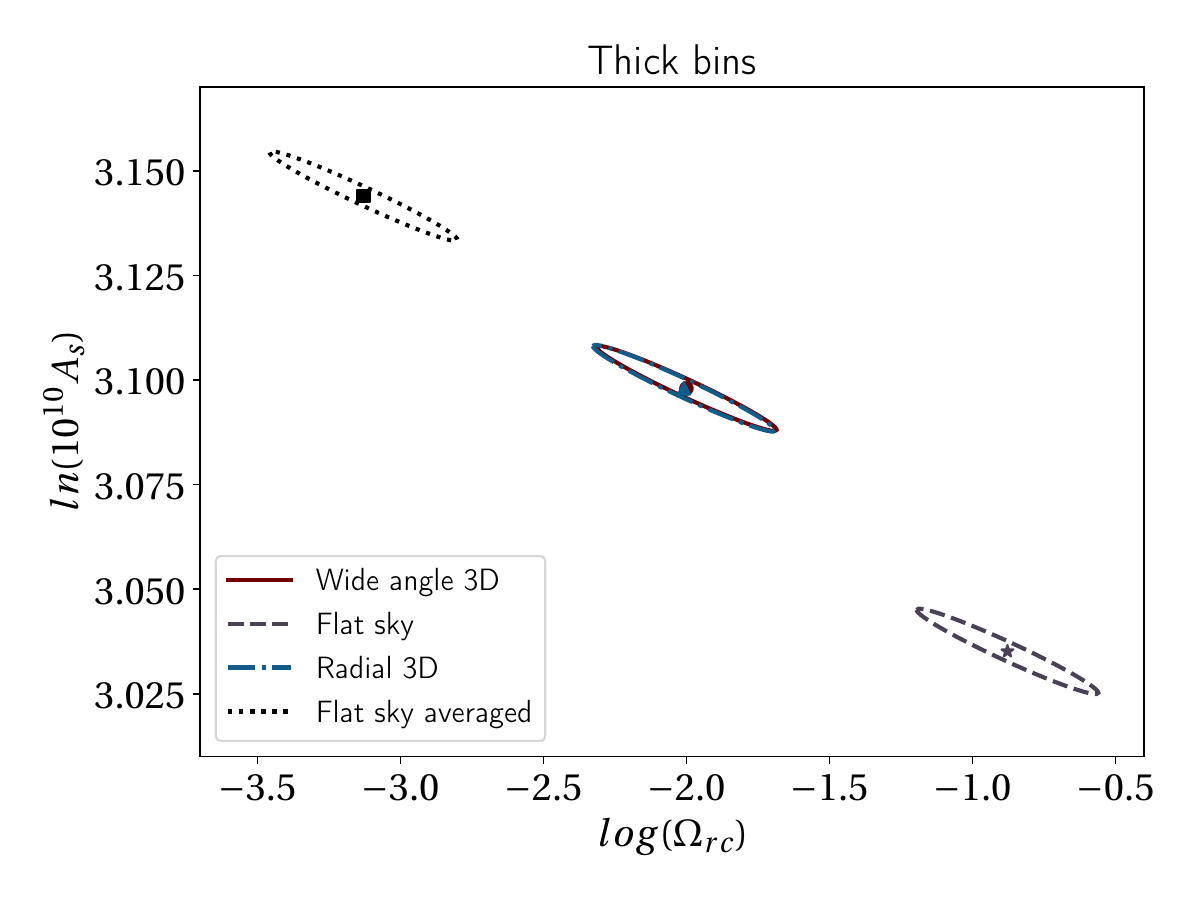}
		\includegraphics[width=.48\textwidth]{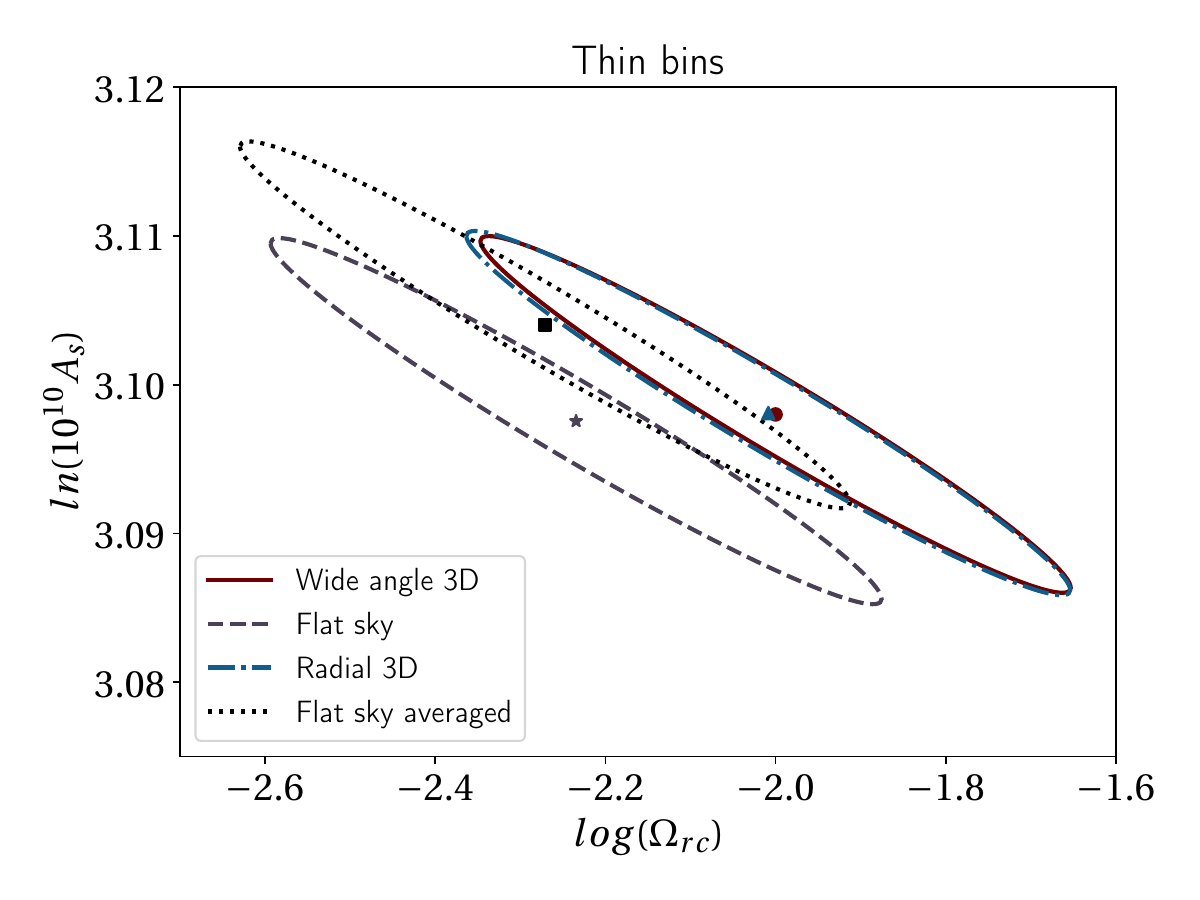}
		\caption{$1\sigma$ constraints and best fit values of $\log\Omega_{\mathrm{rc}}$ and $\ln(10^{10}A_s)$ and in the case of the wide angle 3D model (red continuous line), flat sky averaged (black dotted line), flat sky (grey dashed line) and the radial 3D model (blue dot-dashed line). The 3D radial and the full wide angle 3D cases are basically overlapping in both panels.}
		\label{fig:ellispesndgp}
	\end{figure}

	We perform a similar Fisher analysis of what done in previous sections and check whether the inferred value of the $\Omega_{rc}$ parameter is affected when taking simplifying assumptions in the modeling of the 2PCF.
	As before, we keep the cosmological parameters fixed to their Planck 2018~\cite{Planck2018} values, but we let the primordial amplitude $A_s$ vary, since it is degenerate with $\Omega_{rc}$, as pointed out by~\cite{Piga:2022mge}, using however the Planck prior to mitigate the degeneracy with the bias parameter. Results are showed in Fig.~\ref{fig:ellispesndgp}, where we show the confidence ellipses and the best fit values of the parameters when using the wide angle 3D model, the radial 3D one, the flat sky approximation and the flat sky averaged model.
	
	As for the $\Lambda$CDM case, the results are more relevant for the thick bin case than for the thin bins one.
	When the bins are large, we see that fixing all the sources at the same redshift will induce a relevant shift in the parameters ($\sim3.5\sigma$ for $\log\Omega_{rc}$ and $\sim 6\sigma$ for $\ln(10^{10}A_s)$), while the Radial 3D model assures a similar level of accuracy that of the full wide angle 3D one.

	\section{Conclusions}
	\label{sec:conclusions}
	In this paper we investigated the impact of using approximations in the modeling for galaxy clustering, and how making different assumptions can affect parameter estimation.
	In order to obtain this estimate, we compare the best fit and constraints on two parameters, the growth rate of structures $\gamma$ and the Primordial non-Gaussianity parameter $f_{\rm NL}$, for different modelings of the two-point galaxy correlation function.
	
	We compare results for the full 3D wide angle formalism, which includes all wide angle terms and accounts for galaxies being at different redshifts, with the case where one assumes the plane-parallel and flat sky approximations, hence setting all the galaxies at the same effective redshift and ignoring the aperture angle between galaxies (we call this model Flat sky).
	After this, we consider a model where the redshift distribution of galaxies is accounted for by integrating in the radial direction.
	We perform our calculations for two different survey scenarios, one resembling what will be observed by SPHEREx, and the other for a mock survey with smaller redshift range but thinner bins; we refer to them as ``Thick bins'' and ``Thin bins'' cases, respectively.
	
	Then we introduce a hybrid model, that neglects pure wide angle corrections, but retains the correct modeling in the radial direction; in practice, this means neglecting some of the terms that depend on the aperture angle $\theta$, but keeping terms that use $\theta$ to model the geometry of the system.
	
	We show that using the flat sky model will lead to a considerable shift in the best fit values for both $\gamma$ and $f_{\rm NL}$; such a shift is larger for the thick bins scenario (because of the single effective $z$ used, mis-estimating in this way the growth factor).
	In this case the shift exceeds the 1-$\sigma$ constraints on the parameter, while for the thin bins case the shift is reduced, and will lead to a mis-estimation that stays within the error bars for both parameters.
	In general, the shift induced by such simplifications depend on the thickness of the bins over which the analysis is performed, as fixing the radial position of each source to an effective redshift $z_{\mathrm{eff}}$ is less and less accurate with increasing bin thickness.
	For this fact, measurements of the growth of structure can be significantly affected (up to $3-\sigma$ level for the most extreme cases). 
	For $\fnl$, the shift can be still non negligible but smaller (up to $0.8\sigma$ in the cases we consider), as PNG contributions arising from scale-dependent bias become important at larger scales, where the precision in the measurements is small. Note that a shift on $f_{\rm NL}$ is also induced by neglecting Doppler terms, that mimics an effective non Gaussian behaviour at very large scales. In any case, such shift can induce a mistake in conclusions on inflationary models, especially for future surveys that aim to measure $\sigma _{f_{\rm NL}}<1$.
	In general, error bars will change only marginally.
	
	We argue, however, that even in the cases where the best fit shifts induced by a poor modeling are below the 1-$\sigma$ level, having a consistently biased estimate of some of the parameters can affect more considerably the full parameter estimation process, and therefore such biases need to be avoided.
	
	Finally, we applied our machinery to a modified gravity scenario, namely the nDGP model, in order to investigate the impact of the flat sky approximation on measurements of the model's parameters, showing that also in this case it will lead to errors in their estimates, with similar significance and behaviors as for the $\Lambda$CDM case.
	
	After computing how inaccurate different formalisms are, we tried to find a simplified model that still remains accurate enough.
	Guided by the fact that the main source of induced error seems to come from radial inaccuracies, we focus on the part that preserves the correct modeling in the radial direction.
	
	Hence, we propose a hybrid model, which we call 3D Radial, that neglects the curvature of the sky, but still includes all the terms that are needed to properly model the geometry of the system composed by the observer and the galaxy pairs.
	In practice, we neglect the terms proportional to the opening angle $\theta$ arising from the tripolar basis coefficients in the wide angle expansion.
	Neglecting purely wide-angle terms, the resulting expression is much simpler than the full wide angle case, but this model maintains a high level of accuracy, by way of accounting for the radial position of each galaxy.
	Note however that this is not equivalent to say that we are allowed to use the plane parallel (or distant observer) approximation, as the only way to model the observed position of each pair of galaxies is by describing the system formed by the observer and the two galaxies with 3 degrees of freedom (and not 2 as in the plane parallel case), that in this work are chosen to be $\{s,\theta,\phi\}$.
	
	We show that using this hybrid model the shift in the best fit is greatly reduced for all parameters and cases we consider, therefore showing that this represents a very good approximation that can be safely used for surveys similar to the ones we considered.

	While we will investigate in detail for which cases the curvature of the sky might still be important in the 2PCF modeling in a follow-up paper, our results here are in agreement with recent results showing how radial modes need to be model correctly.
	This is another confirmation that forthcoming galaxy surveys will need to be analyzed with models that are as accurate as possible.

	\section*{Acknowledgments}
	The authors thank Nicola Bellomo, Daniele Bertacca, Camille Bonvin, Mohamed Yousry Elkhashab, Sarah Libanore, Roy Maartens, Federico Semenzato, Eleonora Vanzan and Licia Verde for useful discussions. AR acknowledges funding from the Italian Ministry of University and Research
	(MIUR) through the “Dipartimenti di eccellenza” project “Science of the Universe”.

	\appendix
	\section{$B^{\ell_1,\ell_2,L}_n$ coefficients}\label{app:coeffs}
	In this section we report the complete set of coefficients for the tripolar spherical harmonics expansion:
	
	\begin{equation}
		\begin{aligned}
			\xi(s,\theta,\phi,z_1,z_2)
			&=D(z_1)D(z_2) \sum_{\ell_{1}, \ell_{2}, L, n} B_n^{{\ell_{1} \ell_{2} L}}(z_1,z_2) S_{\ell_{1} \ell_{2} L}(\theta,\phi) \xi_{L}^{n}(s) \; .
		\end{aligned}
	\end{equation}
	The coefficients including redshift space distortions are, as in~\cite{Bertacca:2012tp, Raccanelli:2013gja, Raccanelli:2013dza}:
	\begin{equation}
		\begin{aligned}
			&B_{0}^{000} = (b_1 + \frac{1}{3}f_1)(b_2 + \frac{1}{3}f_2) & &B_{2}^{112} = -\frac{\sqrt{30}}{3}f_1\alpha_1f_2\alpha_2\\
			&B_{1}^{011} = \sqrt{3}(b_1 + \frac{1}{3}f_1) f_2\alpha_2 & &B_{1}^{121} = \frac{2\sqrt{30}}{15}\alpha_1f_2\\
			&B_{0}^{022} = -\frac{2\sqrt{5}}{3} (b_1 + \frac{1}{3}f_1) f_2 & &B_{1}^{213} = -\frac{2\sqrt{105}}{15} f_1f_2\alpha_2\\
			&B_{1}^{101} = -\sqrt{3} f_1\alpha_1 (b_2 + \frac{1}{3}f_2) & &B_{0}^{222} = \frac{4\sqrt{70}}{63} f_1f_2\\
			&B_{2}^{110} = -\frac{\sqrt{3}}{3} f_1\alpha_1f_2\alpha_2 & &B_{1}^{123} = \frac{2\sqrt{105}}{15}\alpha_1f_2\\
			&B_{1}^{211} = -\frac{2\sqrt{30}}{15} f_1f_2\alpha_2 & &B_{0}^{202} = \frac{2\sqrt{5}}{3} f_1 (b_2 + \frac{1}{3}f_2)\\
			&B_{0}^{220} = \frac{4\sqrt{5}}{45} f_1f_2 & &B_{0}^{224} = \frac{4\sqrt{70}}{35} f_1f_2\; .
		\end{aligned}
	\end{equation}
	\section{$F_{L,n}$ corefficients}\label{app:FF}
	In the radial 3D model, the explicit expressions for the $F_{L,n}$ coefficients in Equation~\eqref{eq:radial} are:
	\begin{equation}
		\begin{gathered}   
			F_{00}=\mathcal{F}_1\mathcal{F}_2+\frac{4}{45}f_1 f_2 \, , \\
			F_{20}=-\frac{2}{3}\bigg(\mathcal{F}_1f_2+\mathcal{F}_2f_1\bigg)-\frac{8}{63}f_1f_2 \, , \\
			F_{40}=\frac{8}{35}f_1f_2 \, ,\\
			F_{11}=\alpha_1f_1\mathcal{F}_2-\alpha_2f_2\mathcal{F}_1+\frac{4}{15}f_1f_2\left(\alpha_1-\alpha_2\right) \, ,\\
			F_{31}=\frac{2}{5}f_1f_2(\alpha_2-\alpha_1) \, ,\\
			F_{02}=\frac{\alpha_1\alpha_2f_1f_2}{3} \, ,\\
			F_{22}=-\frac{2\alpha_1\alpha_2f_1f_2}{3}\; ,
		\end{gathered}
	\end{equation}
	where now the $\theta$ dependence of the observable is only encoded in the redshift dependent quantities $\mathcal{F}_i$, $f_i$,$b_i$ and $\alpha_i$ since the redshift of each source can be rewritten as a function of $\{s,\theta,\phi\}$.
	For the non Gaussian terms we have:
	\begin{equation}
		\begin{gathered}
			\tilde{F}_{02}=\Delta b_1\mathcal{F}_2+\Delta b_2\mathcal{F}_1 \\
			\tilde{F}_{04}=\Delta b_1\Delta b_2\\
			\tilde{F}_{22}=-\frac{2}{3}(\Delta b_1f_2+\Delta b_2f_1)\\
			\tilde{F}_{13}=\alpha_2f_2\Delta b_1-\alpha_1f_1\Delta b_2\; ,
		\end{gathered}
	\end{equation}
	where in this case the $\xi^{n}_{L}$ functions in Equation~\eqref{eq:radial} must be replaced by $\xi^{n}_{L,ng}$.
	
	\section{Doppler terms}
	\label{app:doppler}
\begin{figure}
		\centering
		\includegraphics[width=.48\textwidth]{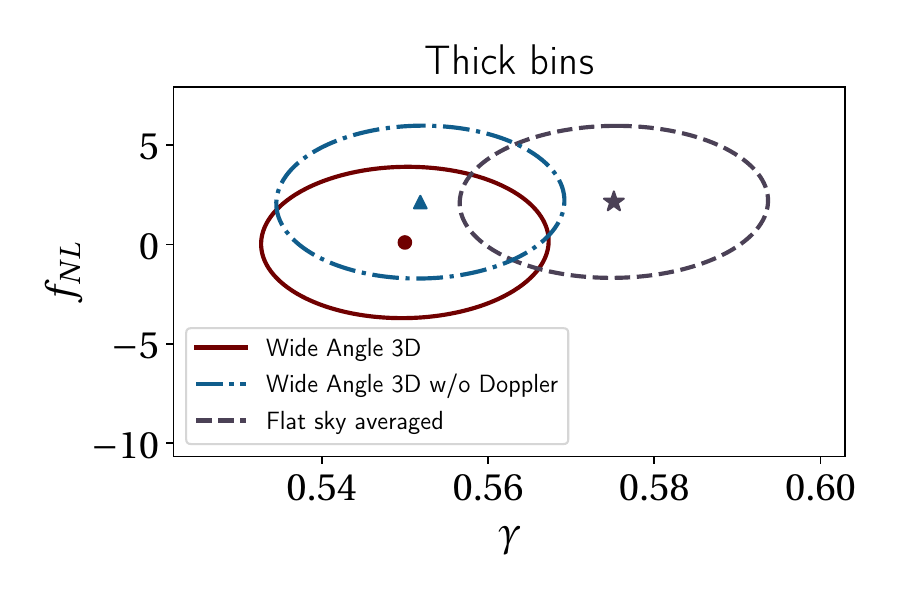}
		\includegraphics[width=.48\textwidth]{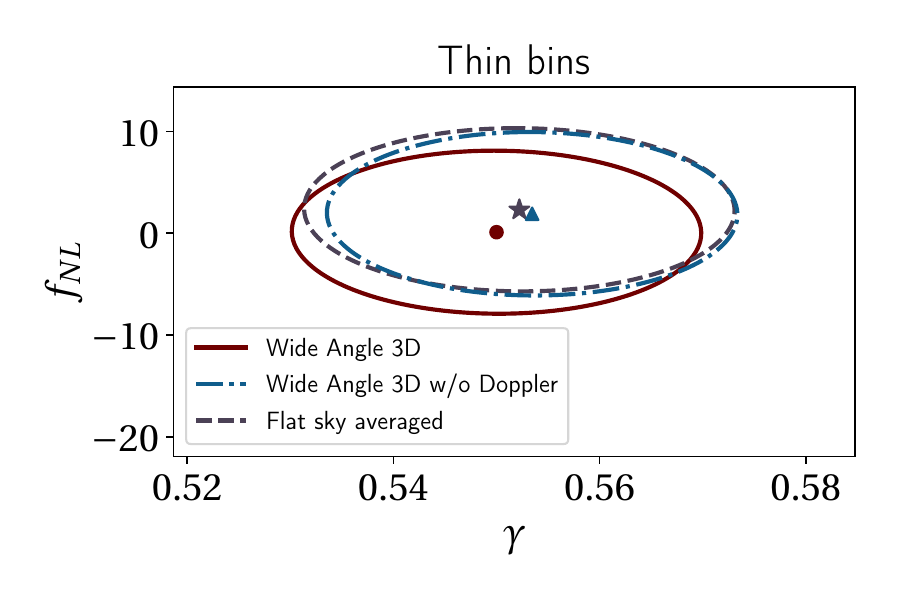}
		\caption{$1\sigma$ constraints and best fit values of $\gamma$ and $\fnl$ in the case of the wide angle 3D model (red continuous line), wide angle 3D model but neglecting Doppler terms (blue dot-dashed line) and flat sky averaged (grey dashed line).}
		\label{fig:ellipses_doppler}
	\end{figure}

	In the full expression of the redshift-space correlation function, the $\alpha$-term is usually called ``Doppler term''; this is a velocity term that is mostly neglected, and does not appear in the standard Kaiser expression (even though it is present in the original paper~\cite{Kaiser1987}). It is suppressed at high-redshift, and is generally subdominant, but it can become relevant in some situations, depending on the selection function and area of the sky surveyed; more details can be found in e.g.,~\cite{Raccanelli:2010hk, Raccanelli:2016avd,Abramo:2017xnp,Elkhashab:2021lsk}. As this term causes effects of so-called ``mode-coupling'', it is responsible for a considerable increase of the computational time required to calculate the 2PCF, and for this reason, here we investigate what is the effect of neglecting it in the modeling. In Fig.~\ref{fig:ellipses_doppler} we show the best fit values and $1\sigma$ constraints for the full wide angle 3D model (solid red), the wide angle 3D model without Doppler terms (blue dot-dashed) and flat sky averaged model (grey dashed). We see that in general the best fit shift induced by Doppler terms remains always $<1\sigma$, for the standard cases we discussed here, and therefore we neglect them for the purposes of this work. However, as mentioned above, the Doppler terms can become relevant, especially for some galaxy populations and in the multi-tracer case (see e.g.,~\cite{Raccanelli:2013dza}); we will investigate in detail these situations and the impact of Doppler terms in our hybrid modeling in a future paper.

	\section{Best fit shift as a function of scale and redshift}
\begin{figure}
		\centering
		\includegraphics[width=.48\textwidth]{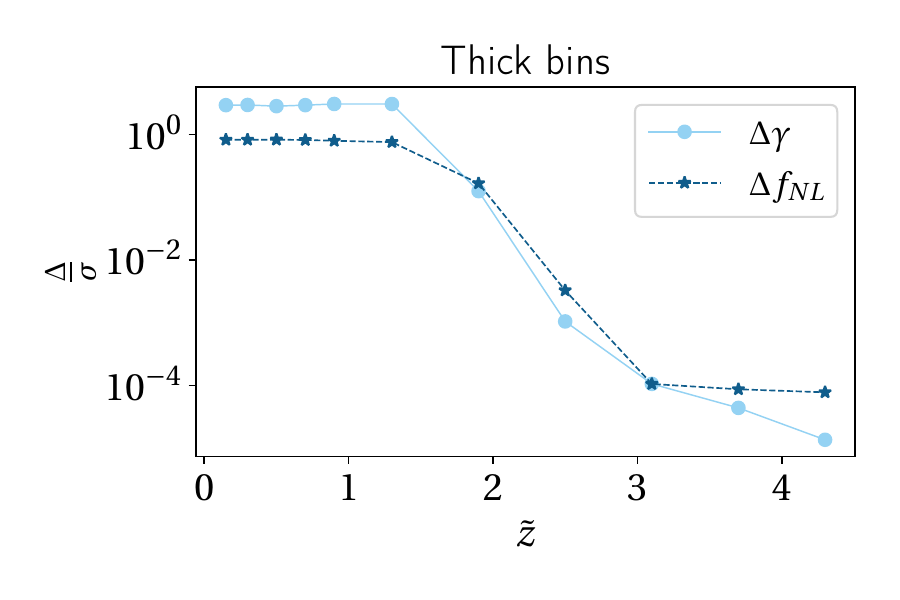}
		\includegraphics[width=.48\textwidth]{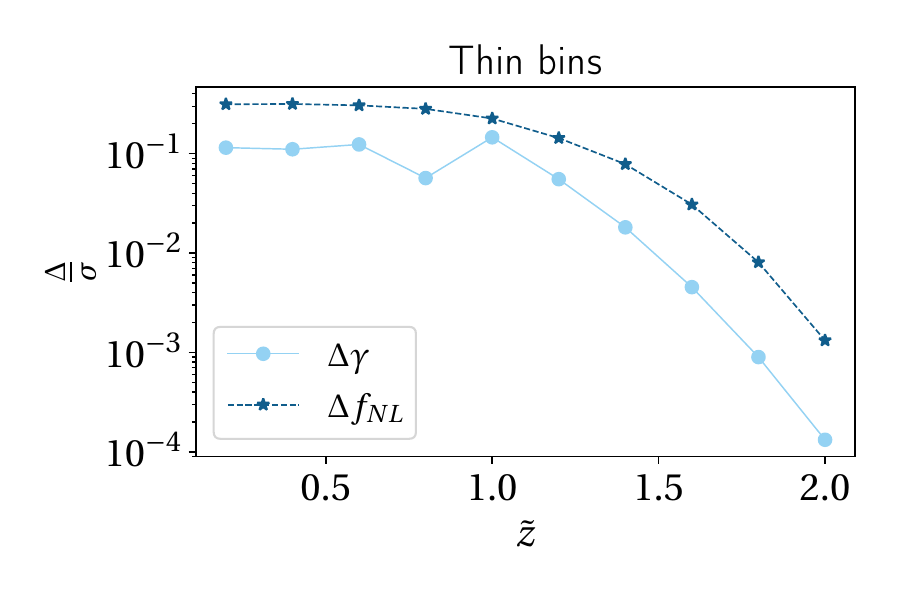}
		
		\caption{Best fit shift of $\gamma$ and $f_{\rm NL}$ obtained if we model with the flat sky limit only bins above $\bar{z}$ .}
		\label{fig:shift_z}
	\end{figure}

	Here, as an additional test, we look at the importance of wide angle 3D corrections as a function of the redshift and scale. To do so, and to give a general guidance for when it is safe to use the standard Kaiser approach, in Fig.~\ref{fig:shift_z} we plot the best fit shifts as a function of the maximum redshift until which we use the full 3D wide angle modeling, which we call $\bar{z}$.
	We see that the shift becomes negligible (and therefore one can use the standard analysis after $\bar{z}$) right after $z\sim 1$, due to the combination of the fact that a large part of the shift due to radial effects comes from the very thick bin at $z=1.3$, and the precision in the measurement decreases as we go to higher redshifts. In the thin bin scenario while the magnitude shift decreases with redshift as the precision in the measurements decreases in general it remains negligible for all redshift as expected.

	\begin{figure}
		\centering
		\includegraphics[width=.48\textwidth]{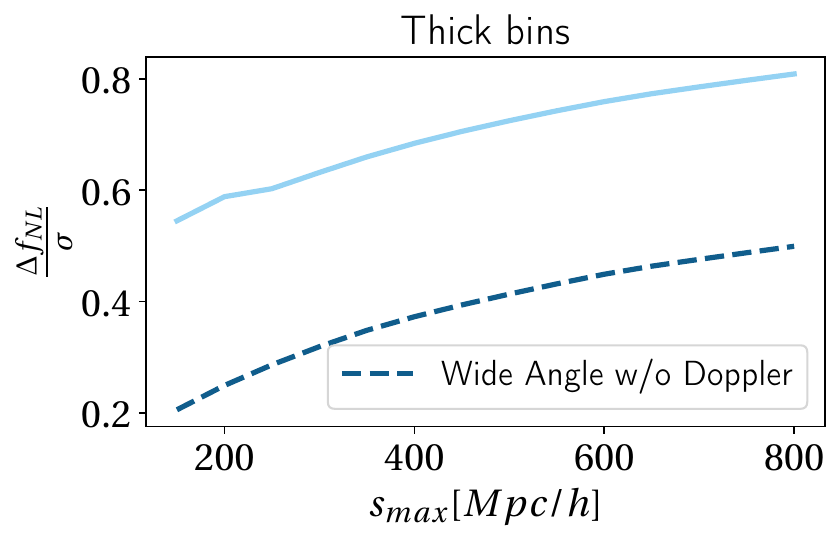}
		\includegraphics[width=.48\textwidth]{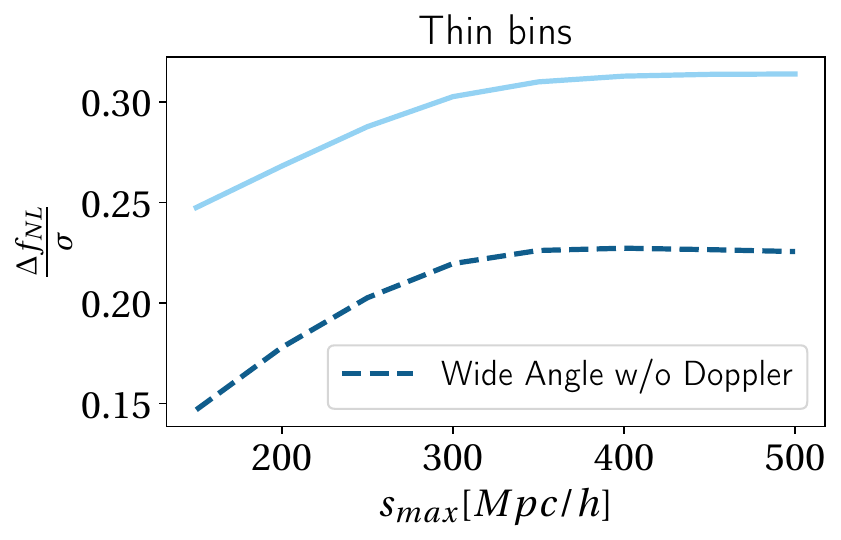}
		
		\caption{Best fit shift of $f_{\rm NL}$ as a function of the maximum scale included in the analysis. The dashed line shows the shift induced by neglecting Doppler terms.}
		\label{fig:shift_s}
	\end{figure}
	In Fig.~\ref{fig:shift_s} we show the best fit shift as function of the maximum scale included in the analysis.
	As PNG contributions from the scale dependent bias are maximized at large scales due to their $k^{-2}$ behaviour, the shift in $f_{\rm NL}$ grows if we analyze very large scales. The shift is due to two contributions: radial effects due to the poor modeling of  the redshift evolution achieved by the flat sky approximation and the effect of Doppler terms that introduce an 'effective PNG' at large scales due to their $k^{-2}$ dependence.
	
	\section*{References}
	\bibliography{ref}

\begin{thebibliography}{116}
\expandafter\ifx\csname natexlab\endcsname\relax\def\natexlab#1{#1}\fi
\expandafter\ifx\csname bibnamefont\endcsname\relax
  \def\bibnamefont#1{#1}\fi
\expandafter\ifx\csname bibfnamefont\endcsname\relax
  \def\bibfnamefont#1{#1}\fi
\expandafter\ifx\csname citenamefont\endcsname\relax
  \def\citenamefont#1{#1}\fi
\expandafter\ifx\csname url\endcsname\relax
  \def\url#1{\texttt{#1}}\fi
\expandafter\ifx\csname urlprefix\endcsname\relax\def\urlprefix{URL }\fi
\providecommand{\bibinfo}[2]{#2}
\providecommand{\eprint}[2][]{\url{#2}}

\bibitem[{\citenamefont{Mellier et~al.}(2024)}]{Euclid:2024yrr}
\bibinfo{author}{\bibfnamefont{Y.}~\bibnamefont{Mellier}} \bibnamefont{et~al.}
  (\bibinfo{collaboration}{Euclid}) (\bibinfo{year}{2024}),
  \eprint{2405.13491}.

\bibitem[{\citenamefont{Laureijs et~al.}(2011)}]{EUCLID:2011zbd}
\bibinfo{author}{\bibfnamefont{R.}~\bibnamefont{Laureijs}} \bibnamefont{et~al.}
  (\bibinfo{collaboration}{EUCLID}) (\bibinfo{year}{2011}), \eprint{1110.3193}.

\bibitem[{\citenamefont{Blanchard et~al.}(2020)}]{Euclid:2019clj}
\bibinfo{author}{\bibfnamefont{A.}~\bibnamefont{Blanchard}}
  \bibnamefont{et~al.} (\bibinfo{collaboration}{Euclid}),
  \bibinfo{journal}{Astron. Astrophys.} \textbf{\bibinfo{volume}{642}},
  \bibinfo{pages}{A191} (\bibinfo{year}{2020}), \eprint{1910.09273}.

\bibitem[{\citenamefont{Aghamousa et~al.}(2016)}]{DESI:2016fyo}
\bibinfo{author}{\bibfnamefont{A.}~\bibnamefont{Aghamousa}}
  \bibnamefont{et~al.} (\bibinfo{collaboration}{DESI}) (\bibinfo{year}{2016}),
  \eprint{1611.00036}.

\bibitem[{\citenamefont{Adame et~al.}(2024)}]{DESI:2024mwx}
\bibinfo{author}{\bibfnamefont{A.~G.} \bibnamefont{Adame}} \bibnamefont{et~al.}
  (\bibinfo{collaboration}{DESI}) (\bibinfo{year}{2024}), \eprint{2404.03002}.

\bibitem[{\citenamefont{Dor\'e et~al.}(2014)}]{Dore:2014cca}
\bibinfo{author}{\bibfnamefont{O.}~\bibnamefont{Dor\'e}} \bibnamefont{et~al.}
  (\bibinfo{year}{2014}), \eprint{1412.4872}.

\bibitem[{\citenamefont{Abate et~al.}(2012)}]{LSSTDarkEnergyScience:2012kar}
\bibinfo{author}{\bibfnamefont{A.}~\bibnamefont{Abate}} \bibnamefont{et~al.}
  (\bibinfo{collaboration}{LSST Dark Energy Science}) (\bibinfo{year}{2012}),
  \eprint{1211.0310}.

\bibitem[{\citenamefont{Ellis et~al.}(2014)}]{PFSTeam:2012fqu}
\bibinfo{author}{\bibfnamefont{R.}~\bibnamefont{Ellis}} \bibnamefont{et~al.}
  (\bibinfo{collaboration}{PFS Team}), \bibinfo{journal}{Publ. Astron. Soc.
  Jap.} \textbf{\bibinfo{volume}{66}}, \bibinfo{pages}{R1}
  (\bibinfo{year}{2014}), \eprint{1206.0737}.

\bibitem[{\citenamefont{Eifler et~al.}(2021)}]{Eifler:2020vvg}
\bibinfo{author}{\bibfnamefont{T.}~\bibnamefont{Eifler}} \bibnamefont{et~al.},
  \bibinfo{journal}{Mon. Not. Roy. Astron. Soc.}
  \textbf{\bibinfo{volume}{507}}, \bibinfo{pages}{1746} (\bibinfo{year}{2021}),
  \eprint{2004.05271}.

\bibitem[{\citenamefont{Wenzl et~al.}(2022)\citenamefont{Wenzl, Doux, Heinrich,
  Bean, Jain, Dor\'e, Eifler, and Fang}}]{Wenzl:2021rrq}
\bibinfo{author}{\bibfnamefont{L.}~\bibnamefont{Wenzl}},
  \bibinfo{author}{\bibfnamefont{C.}~\bibnamefont{Doux}},
  \bibinfo{author}{\bibfnamefont{C.}~\bibnamefont{Heinrich}},
  \bibinfo{author}{\bibfnamefont{R.}~\bibnamefont{Bean}},
  \bibinfo{author}{\bibfnamefont{B.}~\bibnamefont{Jain}},
  \bibinfo{author}{\bibfnamefont{O.}~\bibnamefont{Dor\'e}},
  \bibinfo{author}{\bibfnamefont{T.}~\bibnamefont{Eifler}}, \bibnamefont{and}
  \bibinfo{author}{\bibfnamefont{X.}~\bibnamefont{Fang}},
  \bibinfo{journal}{Mon. Not. Roy. Astron. Soc.}
  \textbf{\bibinfo{volume}{512}}, \bibinfo{pages}{5311} (\bibinfo{year}{2022}),
  \eprint{2112.07681}.

\bibitem[{\citenamefont{Schlegel et~al.}(2022)}]{Schlegel:2022vrv}
\bibinfo{author}{\bibfnamefont{D.~J.} \bibnamefont{Schlegel}}
  \bibnamefont{et~al.} (\bibinfo{year}{2022}), \eprint{2209.04322}.

\bibitem[{\citenamefont{Wang et~al.}(2019)}]{Wang2018}
\bibinfo{author}{\bibfnamefont{Y.}~\bibnamefont{Wang}} \bibnamefont{et~al.},
  \bibinfo{journal}{Publ. Astron. Soc. Austral.} \textbf{\bibinfo{volume}{36}},
  \bibinfo{pages}{e015} (\bibinfo{year}{2019}), \eprint{1802.01539}.

\bibitem[{\citenamefont{{Y.Wang \textit{et al.}, \textit{in prep.}}}()}]{XXX}
\bibinfo{author}{\bibnamefont{{Y.Wang \textit{et al.}, \textit{in prep.}}}}

\bibitem[{\citenamefont{Brieden et~al.}(2020)\citenamefont{Brieden,
  Gil-Mar\'\i{}n, Verde, and Bernal}}]{Brieden:2020upf}
\bibinfo{author}{\bibfnamefont{S.}~\bibnamefont{Brieden}},
  \bibinfo{author}{\bibfnamefont{H.}~\bibnamefont{Gil-Mar\'\i{}n}},
  \bibinfo{author}{\bibfnamefont{L.}~\bibnamefont{Verde}}, \bibnamefont{and}
  \bibinfo{author}{\bibfnamefont{J.~L.} \bibnamefont{Bernal}},
  \bibinfo{journal}{JCAP} \textbf{\bibinfo{volume}{09}}, \bibinfo{pages}{052}
  (\bibinfo{year}{2020}), \eprint{2006.10857}.

\bibitem[{\citenamefont{Kitching et~al.}(2016)\citenamefont{Kitching, Verde,
  Heavens, and Jimenez}}]{Kitching:2016hvn}
\bibinfo{author}{\bibfnamefont{T.~D.} \bibnamefont{Kitching}},
  \bibinfo{author}{\bibfnamefont{L.}~\bibnamefont{Verde}},
  \bibinfo{author}{\bibfnamefont{A.~F.} \bibnamefont{Heavens}},
  \bibnamefont{and} \bibinfo{author}{\bibfnamefont{R.}~\bibnamefont{Jimenez}},
  \bibinfo{journal}{Mon. Not. Roy. Astron. Soc.}
  \textbf{\bibinfo{volume}{459}}, \bibinfo{pages}{971} (\bibinfo{year}{2016}),
  \eprint{1602.02960}.

\bibitem[{\citenamefont{Valcin et~al.}(2021)\citenamefont{Valcin, Jimenez,
  Verde, Bernal, and Wandelt}}]{Valcin:2021jcg}
\bibinfo{author}{\bibfnamefont{D.}~\bibnamefont{Valcin}},
  \bibinfo{author}{\bibfnamefont{R.}~\bibnamefont{Jimenez}},
  \bibinfo{author}{\bibfnamefont{L.}~\bibnamefont{Verde}},
  \bibinfo{author}{\bibfnamefont{J.~L.} \bibnamefont{Bernal}},
  \bibnamefont{and} \bibinfo{author}{\bibfnamefont{B.~D.}
  \bibnamefont{Wandelt}}, \bibinfo{journal}{JCAP}
  \textbf{\bibinfo{volume}{08}}, \bibinfo{pages}{017} (\bibinfo{year}{2021}),
  \eprint{2102.04486}.

\bibitem[{\citenamefont{Norena et~al.}(2012)\citenamefont{Norena, Verde,
  Jimenez, Pena-Garay, and Gomez}}]{Norena:2011sh}
\bibinfo{author}{\bibfnamefont{J.}~\bibnamefont{Norena}},
  \bibinfo{author}{\bibfnamefont{L.}~\bibnamefont{Verde}},
  \bibinfo{author}{\bibfnamefont{R.}~\bibnamefont{Jimenez}},
  \bibinfo{author}{\bibfnamefont{C.}~\bibnamefont{Pena-Garay}},
  \bibnamefont{and} \bibinfo{author}{\bibfnamefont{C.}~\bibnamefont{Gomez}},
  \bibinfo{journal}{Mon. Not. Roy. Astron. Soc.}
  \textbf{\bibinfo{volume}{419}}, \bibinfo{pages}{1040} (\bibinfo{year}{2012}),
  \eprint{1107.0729}.

\bibitem[{\citenamefont{Bernardeau et~al.}(2002)\citenamefont{Bernardeau,
  Colombi, Gaztanaga, and Scoccimarro}}]{Bernardeau:2001qr}
\bibinfo{author}{\bibfnamefont{F.}~\bibnamefont{Bernardeau}},
  \bibinfo{author}{\bibfnamefont{S.}~\bibnamefont{Colombi}},
  \bibinfo{author}{\bibfnamefont{E.}~\bibnamefont{Gaztanaga}},
  \bibnamefont{and}
  \bibinfo{author}{\bibfnamefont{R.}~\bibnamefont{Scoccimarro}},
  \bibinfo{journal}{Phys. Rept.} \textbf{\bibinfo{volume}{367}},
  \bibinfo{pages}{1} (\bibinfo{year}{2002}), \eprint{astro-ph/0112551}.

\bibitem[{\citenamefont{Matarrese and Pietroni}(2007)}]{Matarrese:2007wc}
\bibinfo{author}{\bibfnamefont{S.}~\bibnamefont{Matarrese}} \bibnamefont{and}
  \bibinfo{author}{\bibfnamefont{M.}~\bibnamefont{Pietroni}},
  \bibinfo{journal}{JCAP} \textbf{\bibinfo{volume}{06}}, \bibinfo{pages}{026}
  (\bibinfo{year}{2007}), \eprint{astro-ph/0703563}.

\bibitem[{\citenamefont{Pietroni}(2008)}]{Pietroni:2008jx}
\bibinfo{author}{\bibfnamefont{M.}~\bibnamefont{Pietroni}},
  \bibinfo{journal}{JCAP} \textbf{\bibinfo{volume}{10}}, \bibinfo{pages}{036}
  (\bibinfo{year}{2008}), \eprint{0806.0971}.

\bibitem[{\citenamefont{Matsubara}(2008{\natexlab{a}})}]{Matsubara:2007wj}
\bibinfo{author}{\bibfnamefont{T.}~\bibnamefont{Matsubara}},
  \bibinfo{journal}{Phys. Rev. D} \textbf{\bibinfo{volume}{77}},
  \bibinfo{pages}{063530} (\bibinfo{year}{2008}{\natexlab{a}}),
  \eprint{0711.2521}.

\bibitem[{\citenamefont{Matsubara}(2008{\natexlab{b}})}]{Matsubara:2008wx}
\bibinfo{author}{\bibfnamefont{T.}~\bibnamefont{Matsubara}},
  \bibinfo{journal}{Phys. Rev. D} \textbf{\bibinfo{volume}{78}},
  \bibinfo{pages}{083519} (\bibinfo{year}{2008}{\natexlab{b}}),
  \bibinfo{note}{[Erratum: Phys.Rev.D 78, 109901 (2008)]}, \eprint{0807.1733}.

\bibitem[{\citenamefont{Takahashi et~al.}(2012)\citenamefont{Takahashi, Sato,
  Nishimichi, Taruya, and Oguri}}]{Takahashi:2012em}
\bibinfo{author}{\bibfnamefont{R.}~\bibnamefont{Takahashi}},
  \bibinfo{author}{\bibfnamefont{M.}~\bibnamefont{Sato}},
  \bibinfo{author}{\bibfnamefont{T.}~\bibnamefont{Nishimichi}},
  \bibinfo{author}{\bibfnamefont{A.}~\bibnamefont{Taruya}}, \bibnamefont{and}
  \bibinfo{author}{\bibfnamefont{M.}~\bibnamefont{Oguri}},
  \bibinfo{journal}{Astrophys. J.} \textbf{\bibinfo{volume}{761}},
  \bibinfo{pages}{152} (\bibinfo{year}{2012}), \eprint{1208.2701}.

\bibitem[{\citenamefont{Chen et~al.}(2020)\citenamefont{Chen, Vlah, and
  White}}]{Chen:2020fxs}
\bibinfo{author}{\bibfnamefont{S.-F.} \bibnamefont{Chen}},
  \bibinfo{author}{\bibfnamefont{Z.}~\bibnamefont{Vlah}}, \bibnamefont{and}
  \bibinfo{author}{\bibfnamefont{M.}~\bibnamefont{White}},
  \bibinfo{journal}{JCAP} \textbf{\bibinfo{volume}{07}}, \bibinfo{pages}{062}
  (\bibinfo{year}{2020}), \eprint{2005.00523}.

\bibitem[{\citenamefont{Reid and White}(2011)}]{Reid:2011ar}
\bibinfo{author}{\bibfnamefont{B.~A.} \bibnamefont{Reid}} \bibnamefont{and}
  \bibinfo{author}{\bibfnamefont{M.}~\bibnamefont{White}},
  \bibinfo{journal}{Mon. Not. Roy. Astron. Soc.}
  \textbf{\bibinfo{volume}{417}}, \bibinfo{pages}{1913} (\bibinfo{year}{2011}),
  \eprint{1105.4165}.

\bibitem[{\citenamefont{Vlah and White}(2019)}]{Vlah:2018ygt}
\bibinfo{author}{\bibfnamefont{Z.}~\bibnamefont{Vlah}} \bibnamefont{and}
  \bibinfo{author}{\bibfnamefont{M.}~\bibnamefont{White}},
  \bibinfo{journal}{JCAP} \textbf{\bibinfo{volume}{03}}, \bibinfo{pages}{007}
  (\bibinfo{year}{2019}), \eprint{1812.02775}.

\bibitem[{\citenamefont{Vlah et~al.}(2016)\citenamefont{Vlah, Seljak, Chu, and
  Feng}}]{Vlah:2015zda}
\bibinfo{author}{\bibfnamefont{Z.}~\bibnamefont{Vlah}},
  \bibinfo{author}{\bibfnamefont{U.}~\bibnamefont{Seljak}},
  \bibinfo{author}{\bibfnamefont{M.~Y.} \bibnamefont{Chu}}, \bibnamefont{and}
  \bibinfo{author}{\bibfnamefont{Y.}~\bibnamefont{Feng}},
  \bibinfo{journal}{JCAP} \textbf{\bibinfo{volume}{03}}, \bibinfo{pages}{057}
  (\bibinfo{year}{2016}), \eprint{1509.02120}.

\bibitem[{\citenamefont{Vlah et~al.}(2015)\citenamefont{Vlah, White, and
  Aviles}}]{Vlah:2015sea}
\bibinfo{author}{\bibfnamefont{Z.}~\bibnamefont{Vlah}},
  \bibinfo{author}{\bibfnamefont{M.}~\bibnamefont{White}}, \bibnamefont{and}
  \bibinfo{author}{\bibfnamefont{A.}~\bibnamefont{Aviles}},
  \bibinfo{journal}{JCAP} \textbf{\bibinfo{volume}{09}}, \bibinfo{pages}{014}
  (\bibinfo{year}{2015}), \eprint{1506.05264}.

\bibitem[{\citenamefont{D'Amico et~al.}(2021)\citenamefont{D'Amico, Marinucci,
  Pietroni, and Vernizzi}}]{DAmico:2021rdb}
\bibinfo{author}{\bibfnamefont{G.}~\bibnamefont{D'Amico}},
  \bibinfo{author}{\bibfnamefont{M.}~\bibnamefont{Marinucci}},
  \bibinfo{author}{\bibfnamefont{M.}~\bibnamefont{Pietroni}}, \bibnamefont{and}
  \bibinfo{author}{\bibfnamefont{F.}~\bibnamefont{Vernizzi}},
  \bibinfo{journal}{JCAP} \textbf{\bibinfo{volume}{10}}, \bibinfo{pages}{069}
  (\bibinfo{year}{2021}), \eprint{2109.09573}.

\bibitem[{\citenamefont{Fasiello et~al.}(2022)\citenamefont{Fasiello, Fujita,
  and Vlah}}]{Fasiello:2022lff}
\bibinfo{author}{\bibfnamefont{M.}~\bibnamefont{Fasiello}},
  \bibinfo{author}{\bibfnamefont{T.}~\bibnamefont{Fujita}}, \bibnamefont{and}
  \bibinfo{author}{\bibfnamefont{Z.}~\bibnamefont{Vlah}},
  \bibinfo{journal}{Phys. Rev. D} \textbf{\bibinfo{volume}{106}},
  \bibinfo{pages}{123504} (\bibinfo{year}{2022}), \eprint{2205.10026}.

\bibitem[{\citenamefont{Senatore and Zaldarriaga}(2014)}]{Senatore:2014vja}
\bibinfo{author}{\bibfnamefont{L.}~\bibnamefont{Senatore}} \bibnamefont{and}
  \bibinfo{author}{\bibfnamefont{M.}~\bibnamefont{Zaldarriaga}}
  (\bibinfo{year}{2014}), \eprint{1409.1225}.

\bibitem[{\citenamefont{Perko et~al.}(2016)\citenamefont{Perko, Senatore,
  Jennings, and Wechsler}}]{Perko:2016puo}
\bibinfo{author}{\bibfnamefont{A.}~\bibnamefont{Perko}},
  \bibinfo{author}{\bibfnamefont{L.}~\bibnamefont{Senatore}},
  \bibinfo{author}{\bibfnamefont{E.}~\bibnamefont{Jennings}}, \bibnamefont{and}
  \bibinfo{author}{\bibfnamefont{R.~H.} \bibnamefont{Wechsler}}
  (\bibinfo{year}{2016}), \eprint{1610.09321}.

\bibitem[{\citenamefont{Porto et~al.}(2014)\citenamefont{Porto, Senatore, and
  Zaldarriaga}}]{Porto:2013qua}
\bibinfo{author}{\bibfnamefont{R.~A.} \bibnamefont{Porto}},
  \bibinfo{author}{\bibfnamefont{L.}~\bibnamefont{Senatore}}, \bibnamefont{and}
  \bibinfo{author}{\bibfnamefont{M.}~\bibnamefont{Zaldarriaga}},
  \bibinfo{journal}{JCAP} \textbf{\bibinfo{volume}{05}}, \bibinfo{pages}{022}
  (\bibinfo{year}{2014}), \eprint{1311.2168}.

\bibitem[{\citenamefont{Carrasco et~al.}(2012)\citenamefont{Carrasco,
  Hertzberg, and Senatore}}]{Carrasco:2012cv}
\bibinfo{author}{\bibfnamefont{J.~J.~M.} \bibnamefont{Carrasco}},
  \bibinfo{author}{\bibfnamefont{M.~P.} \bibnamefont{Hertzberg}},
  \bibnamefont{and} \bibinfo{author}{\bibfnamefont{L.}~\bibnamefont{Senatore}},
  \bibinfo{journal}{JHEP} \textbf{\bibinfo{volume}{09}}, \bibinfo{pages}{082}
  (\bibinfo{year}{2012}), \eprint{1206.2926}.

\bibitem[{\citenamefont{Baumann et~al.}(2012)\citenamefont{Baumann, Nicolis,
  Senatore, and Zaldarriaga}}]{Baumann:2010tm}
\bibinfo{author}{\bibfnamefont{D.}~\bibnamefont{Baumann}},
  \bibinfo{author}{\bibfnamefont{A.}~\bibnamefont{Nicolis}},
  \bibinfo{author}{\bibfnamefont{L.}~\bibnamefont{Senatore}}, \bibnamefont{and}
  \bibinfo{author}{\bibfnamefont{M.}~\bibnamefont{Zaldarriaga}},
  \bibinfo{journal}{JCAP} \textbf{\bibinfo{volume}{07}}, \bibinfo{pages}{051}
  (\bibinfo{year}{2012}), \eprint{1004.2488}.

\bibitem[{\citenamefont{Wang et~al.}(2023)\citenamefont{Wang, Jeong, Taruya,
  Nishimichi, and Osato}}]{Wang:2022itv}
\bibinfo{author}{\bibfnamefont{Z.}~\bibnamefont{Wang}},
  \bibinfo{author}{\bibfnamefont{D.}~\bibnamefont{Jeong}},
  \bibinfo{author}{\bibfnamefont{A.}~\bibnamefont{Taruya}},
  \bibinfo{author}{\bibfnamefont{T.}~\bibnamefont{Nishimichi}},
  \bibnamefont{and} \bibinfo{author}{\bibfnamefont{K.}~\bibnamefont{Osato}},
  \bibinfo{journal}{Phys. Rev. D} \textbf{\bibinfo{volume}{107}},
  \bibinfo{pages}{103534} (\bibinfo{year}{2023}), \eprint{2209.00033}.

\bibitem[{\citenamefont{Philcox and Eisenstein}(2020)}]{Philcox:2019hdi}
\bibinfo{author}{\bibfnamefont{O.~H.~E.} \bibnamefont{Philcox}}
  \bibnamefont{and} \bibinfo{author}{\bibfnamefont{D.~J.}
  \bibnamefont{Eisenstein}}, \bibinfo{journal}{Mon. Not. Roy. Astron. Soc.}
  \textbf{\bibinfo{volume}{492}}, \bibinfo{pages}{1214} (\bibinfo{year}{2020}),
  \eprint{1912.01010}.

\bibitem[{\citenamefont{Taruya et~al.}(2010)\citenamefont{Taruya, Nishimichi,
  and Saito}}]{Taruya:2010mx}
\bibinfo{author}{\bibfnamefont{A.}~\bibnamefont{Taruya}},
  \bibinfo{author}{\bibfnamefont{T.}~\bibnamefont{Nishimichi}},
  \bibnamefont{and} \bibinfo{author}{\bibfnamefont{S.}~\bibnamefont{Saito}},
  \bibinfo{journal}{Phys. Rev. D} \textbf{\bibinfo{volume}{82}},
  \bibinfo{pages}{063522} (\bibinfo{year}{2010}), \eprint{1006.0699}.

\bibitem[{\citenamefont{Scoccimarro}(2004)}]{Scoccimarro:2004tg}
\bibinfo{author}{\bibfnamefont{R.}~\bibnamefont{Scoccimarro}},
  \bibinfo{journal}{Phys. Rev. D} \textbf{\bibinfo{volume}{70}},
  \bibinfo{pages}{083007} (\bibinfo{year}{2004}), \eprint{astro-ph/0407214}.

\bibitem[{\citenamefont{de~la Torre and Guzzo}(2012)}]{delaTorre:2012dg}
\bibinfo{author}{\bibfnamefont{S.}~\bibnamefont{de~la Torre}} \bibnamefont{and}
  \bibinfo{author}{\bibfnamefont{L.}~\bibnamefont{Guzzo}},
  \bibinfo{journal}{Mon. Not. Roy. Astron. Soc.}
  \textbf{\bibinfo{volume}{427}}, \bibinfo{pages}{327} (\bibinfo{year}{2012}),
  \eprint{1202.5559}.

\bibitem[{\citenamefont{Szalay et~al.}(1998)\citenamefont{Szalay, Matsubara,
  and Landy}}]{Szalay:1997cc}
\bibinfo{author}{\bibfnamefont{A.~S.} \bibnamefont{Szalay}},
  \bibinfo{author}{\bibfnamefont{T.}~\bibnamefont{Matsubara}},
  \bibnamefont{and} \bibinfo{author}{\bibfnamefont{S.~D.} \bibnamefont{Landy}},
  \bibinfo{journal}{Astrophys. J. Lett.} \textbf{\bibinfo{volume}{498}},
  \bibinfo{pages}{L1} (\bibinfo{year}{1998}), \eprint{astro-ph/9712007}.

\bibitem[{\citenamefont{Bertacca et~al.}(2012)\citenamefont{Bertacca, Maartens,
  Raccanelli, and Clarkson}}]{Bertacca:2012tp}
\bibinfo{author}{\bibfnamefont{D.}~\bibnamefont{Bertacca}},
  \bibinfo{author}{\bibfnamefont{R.}~\bibnamefont{Maartens}},
  \bibinfo{author}{\bibfnamefont{A.}~\bibnamefont{Raccanelli}},
  \bibnamefont{and} \bibinfo{author}{\bibfnamefont{C.}~\bibnamefont{Clarkson}},
  \bibinfo{journal}{JCAP} \textbf{\bibinfo{volume}{10}}, \bibinfo{pages}{025}
  (\bibinfo{year}{2012}), \eprint{1205.5221}.

\bibitem[{\citenamefont{Bertacca et~al.}(2018)\citenamefont{Bertacca,
  Raccanelli, Bartolo, Liguori, Matarrese, and Verde}}]{Bertacca:2017dzm}
\bibinfo{author}{\bibfnamefont{D.}~\bibnamefont{Bertacca}},
  \bibinfo{author}{\bibfnamefont{A.}~\bibnamefont{Raccanelli}},
  \bibinfo{author}{\bibfnamefont{N.}~\bibnamefont{Bartolo}},
  \bibinfo{author}{\bibfnamefont{M.}~\bibnamefont{Liguori}},
  \bibinfo{author}{\bibfnamefont{S.}~\bibnamefont{Matarrese}},
  \bibnamefont{and} \bibinfo{author}{\bibfnamefont{L.}~\bibnamefont{Verde}},
  \bibinfo{journal}{Phys. Rev. D} \textbf{\bibinfo{volume}{97}},
  \bibinfo{pages}{023531} (\bibinfo{year}{2018}), \eprint{1705.09306}.

\bibitem[{\citenamefont{Bertacca}(2020)}]{Bertacca:2019wyg}
\bibinfo{author}{\bibfnamefont{D.}~\bibnamefont{Bertacca}},
  \bibinfo{journal}{Int. J. Mod. Phys. D} \textbf{\bibinfo{volume}{29}},
  \bibinfo{pages}{2050085} (\bibinfo{year}{2020}), \eprint{1912.06887}.

\bibitem[{\citenamefont{Raccanelli et~al.}(2010)\citenamefont{Raccanelli,
  Samushia, and Percival}}]{Raccanelli:2010hk}
\bibinfo{author}{\bibfnamefont{A.}~\bibnamefont{Raccanelli}},
  \bibinfo{author}{\bibfnamefont{L.}~\bibnamefont{Samushia}}, \bibnamefont{and}
  \bibinfo{author}{\bibfnamefont{W.~J.} \bibnamefont{Percival}},
  \bibinfo{journal}{Mon. Not. Roy. Astron. Soc.}
  \textbf{\bibinfo{volume}{409}}, \bibinfo{pages}{1525} (\bibinfo{year}{2010}),
  \eprint{1006.1652}.

\bibitem[{\citenamefont{Raccanelli et~al.}(2013)\citenamefont{Raccanelli,
  Bertacca, Pietrobon, Schmidt, Samushia, Bartolo, Dore, Matarrese, and
  Percival}}]{Raccanelli:2012gt}
\bibinfo{author}{\bibfnamefont{A.}~\bibnamefont{Raccanelli}},
  \bibinfo{author}{\bibfnamefont{D.}~\bibnamefont{Bertacca}},
  \bibinfo{author}{\bibfnamefont{D.}~\bibnamefont{Pietrobon}},
  \bibinfo{author}{\bibfnamefont{F.}~\bibnamefont{Schmidt}},
  \bibinfo{author}{\bibfnamefont{L.}~\bibnamefont{Samushia}},
  \bibinfo{author}{\bibfnamefont{N.}~\bibnamefont{Bartolo}},
  \bibinfo{author}{\bibfnamefont{O.}~\bibnamefont{Dore}},
  \bibinfo{author}{\bibfnamefont{S.}~\bibnamefont{Matarrese}},
  \bibnamefont{and} \bibinfo{author}{\bibfnamefont{W.~J.}
  \bibnamefont{Percival}}, \bibinfo{journal}{Mon. Not. Roy. Astron. Soc.}
  \textbf{\bibinfo{volume}{436}}, \bibinfo{pages}{89} (\bibinfo{year}{2013}),
  \eprint{1207.0500}.

\bibitem[{\citenamefont{Raccanelli et~al.}(2014)\citenamefont{Raccanelli,
  Bertacca, Dor\'e, and Maartens}}]{Raccanelli:2013dza}
\bibinfo{author}{\bibfnamefont{A.}~\bibnamefont{Raccanelli}},
  \bibinfo{author}{\bibfnamefont{D.}~\bibnamefont{Bertacca}},
  \bibinfo{author}{\bibfnamefont{O.}~\bibnamefont{Dor\'e}}, \bibnamefont{and}
  \bibinfo{author}{\bibfnamefont{R.}~\bibnamefont{Maartens}},
  \bibinfo{journal}{JCAP} \textbf{\bibinfo{volume}{08}}, \bibinfo{pages}{022}
  (\bibinfo{year}{2014}), \eprint{1306.6646}.

\bibitem[{\citenamefont{Raccanelli
  et~al.}(2016{\natexlab{a}})\citenamefont{Raccanelli, Bertacca, Maartens,
  Clarkson, and Dor\'e}}]{Raccanelli:2013gja}
\bibinfo{author}{\bibfnamefont{A.}~\bibnamefont{Raccanelli}},
  \bibinfo{author}{\bibfnamefont{D.}~\bibnamefont{Bertacca}},
  \bibinfo{author}{\bibfnamefont{R.}~\bibnamefont{Maartens}},
  \bibinfo{author}{\bibfnamefont{C.}~\bibnamefont{Clarkson}}, \bibnamefont{and}
  \bibinfo{author}{\bibfnamefont{O.}~\bibnamefont{Dor\'e}},
  \bibinfo{journal}{Gen. Rel. Grav.} \textbf{\bibinfo{volume}{48}},
  \bibinfo{pages}{84} (\bibinfo{year}{2016}{\natexlab{a}}), \eprint{1311.6813}.

\bibitem[{\citenamefont{Raccanelli
  et~al.}(2016{\natexlab{b}})\citenamefont{Raccanelli, Montanari, Bertacca,
  Dor\'e, and Durrer}}]{Raccanelli:2015vla}
\bibinfo{author}{\bibfnamefont{A.}~\bibnamefont{Raccanelli}},
  \bibinfo{author}{\bibfnamefont{F.}~\bibnamefont{Montanari}},
  \bibinfo{author}{\bibfnamefont{D.}~\bibnamefont{Bertacca}},
  \bibinfo{author}{\bibfnamefont{O.}~\bibnamefont{Dor\'e}}, \bibnamefont{and}
  \bibinfo{author}{\bibfnamefont{R.}~\bibnamefont{Durrer}},
  \bibinfo{journal}{JCAP} \textbf{\bibinfo{volume}{05}}, \bibinfo{pages}{009}
  (\bibinfo{year}{2016}{\natexlab{b}}), \eprint{1505.06179}.

\bibitem[{\citenamefont{Raccanelli et~al.}(2018)\citenamefont{Raccanelli,
  Bertacca, Jeong, Neyrinck, and Szalay}}]{Raccanelli:2016avd}
\bibinfo{author}{\bibfnamefont{A.}~\bibnamefont{Raccanelli}},
  \bibinfo{author}{\bibfnamefont{D.}~\bibnamefont{Bertacca}},
  \bibinfo{author}{\bibfnamefont{D.}~\bibnamefont{Jeong}},
  \bibinfo{author}{\bibfnamefont{M.~C.} \bibnamefont{Neyrinck}},
  \bibnamefont{and} \bibinfo{author}{\bibfnamefont{A.~S.}
  \bibnamefont{Szalay}}, \bibinfo{journal}{Phys. Dark Univ.}
  \textbf{\bibinfo{volume}{19}}, \bibinfo{pages}{109} (\bibinfo{year}{2018}),
  \eprint{1602.03186}.

\bibitem[{\citenamefont{Raccanelli and
  Vlah}(2023{\natexlab{a}})}]{Raccanelli:2023fle}
\bibinfo{author}{\bibfnamefont{A.}~\bibnamefont{Raccanelli}} \bibnamefont{and}
  \bibinfo{author}{\bibfnamefont{Z.}~\bibnamefont{Vlah}}
  (\bibinfo{year}{2023}{\natexlab{a}}), \eprint{2305.16278}.

\bibitem[{\citenamefont{Raccanelli and
  Vlah}(2023{\natexlab{b}})}]{Raccanelli:2023zkj}
\bibinfo{author}{\bibfnamefont{A.}~\bibnamefont{Raccanelli}} \bibnamefont{and}
  \bibinfo{author}{\bibfnamefont{Z.}~\bibnamefont{Vlah}},
  \bibinfo{journal}{Phys. Rev. D} \textbf{\bibinfo{volume}{108}},
  \bibinfo{pages}{043537} (\bibinfo{year}{2023}{\natexlab{b}}),
  \eprint{2306.00808}.

\bibitem[{\citenamefont{Elkhashab et~al.}(2021)\citenamefont{Elkhashab,
  Porciani, and Bertacca}}]{Elkhashab:2021lsk}
\bibinfo{author}{\bibfnamefont{M.~Y.} \bibnamefont{Elkhashab}},
  \bibinfo{author}{\bibfnamefont{C.}~\bibnamefont{Porciani}}, \bibnamefont{and}
  \bibinfo{author}{\bibfnamefont{D.}~\bibnamefont{Bertacca}},
  \bibinfo{journal}{Mon. Not. Roy. Astron. Soc.}
  \textbf{\bibinfo{volume}{509}}, \bibinfo{pages}{1626} (\bibinfo{year}{2021}),
  \eprint{2108.13424}.

\bibitem[{\citenamefont{Papai and Szapudi}(2008)}]{Papai:2008bd}
\bibinfo{author}{\bibfnamefont{P.}~\bibnamefont{Papai}} \bibnamefont{and}
  \bibinfo{author}{\bibfnamefont{I.}~\bibnamefont{Szapudi}},
  \bibinfo{journal}{Mon. Not. Roy. Astron. Soc.}
  \textbf{\bibinfo{volume}{389}}, \bibinfo{pages}{292} (\bibinfo{year}{2008}),
  \eprint{0802.2940}.

\bibitem[{\citenamefont{Matsubara}(2000)}]{Matsubara:1999du}
\bibinfo{author}{\bibfnamefont{T.}~\bibnamefont{Matsubara}},
  \bibinfo{journal}{Astrophys. J.} \textbf{\bibinfo{volume}{535}},
  \bibinfo{pages}{1} (\bibinfo{year}{2000}), \eprint{astro-ph/9908056}.

\bibitem[{\citenamefont{Zaroubi and Hoffman}(1993)}]{Zaroubi:1993qt}
\bibinfo{author}{\bibfnamefont{S.}~\bibnamefont{Zaroubi}} \bibnamefont{and}
  \bibinfo{author}{\bibfnamefont{Y.}~\bibnamefont{Hoffman}}
  (\bibinfo{year}{1993}), \eprint{astro-ph/9311013}.

\bibitem[{\citenamefont{Di~Dio et~al.}(2014)\citenamefont{Di~Dio, Montanari,
  Durrer, and Lesgourgues}}]{DiDio:2013sea}
\bibinfo{author}{\bibfnamefont{E.}~\bibnamefont{Di~Dio}},
  \bibinfo{author}{\bibfnamefont{F.}~\bibnamefont{Montanari}},
  \bibinfo{author}{\bibfnamefont{R.}~\bibnamefont{Durrer}}, \bibnamefont{and}
  \bibinfo{author}{\bibfnamefont{J.}~\bibnamefont{Lesgourgues}},
  \bibinfo{journal}{JCAP} \textbf{\bibinfo{volume}{01}}, \bibinfo{pages}{042}
  (\bibinfo{year}{2014}), \eprint{1308.6186}.

\bibitem[{\citenamefont{Bonvin and Durrer}(2011)}]{Bonvin:2011bg}
\bibinfo{author}{\bibfnamefont{C.}~\bibnamefont{Bonvin}} \bibnamefont{and}
  \bibinfo{author}{\bibfnamefont{R.}~\bibnamefont{Durrer}},
  \bibinfo{journal}{Phys. Rev. D} \textbf{\bibinfo{volume}{84}},
  \bibinfo{pages}{063505} (\bibinfo{year}{2011}), \eprint{1105.5280}.

\bibitem[{\citenamefont{Castorina and White}(2018)}]{Castorina:2017inr}
\bibinfo{author}{\bibfnamefont{E.}~\bibnamefont{Castorina}} \bibnamefont{and}
  \bibinfo{author}{\bibfnamefont{M.}~\bibnamefont{White}},
  \bibinfo{journal}{Mon. Not. Roy. Astron. Soc.}
  \textbf{\bibinfo{volume}{476}}, \bibinfo{pages}{4403} (\bibinfo{year}{2018}),
  \eprint{1709.09730}.

\bibitem[{\citenamefont{Castorina and White}(2020)}]{Castorina:2019hyr}
\bibinfo{author}{\bibfnamefont{E.}~\bibnamefont{Castorina}} \bibnamefont{and}
  \bibinfo{author}{\bibfnamefont{M.}~\bibnamefont{White}},
  \bibinfo{journal}{Mon. Not. Roy. Astron. Soc.}
  \textbf{\bibinfo{volume}{499}}, \bibinfo{pages}{893} (\bibinfo{year}{2020}),
  \eprint{1911.08353}.

\bibitem[{\citenamefont{Jeong et~al.}(2012)\citenamefont{Jeong, Schmidt, and
  Hirata}}]{Jeong:2011as}
\bibinfo{author}{\bibfnamefont{D.}~\bibnamefont{Jeong}},
  \bibinfo{author}{\bibfnamefont{F.}~\bibnamefont{Schmidt}}, \bibnamefont{and}
  \bibinfo{author}{\bibfnamefont{C.~M.} \bibnamefont{Hirata}},
  \bibinfo{journal}{Phys. Rev. D} \textbf{\bibinfo{volume}{85}},
  \bibinfo{pages}{023504} (\bibinfo{year}{2012}), \eprint{1107.5427}.

\bibitem[{\citenamefont{Yoo}(2014)}]{Yoo:2014kpa}
\bibinfo{author}{\bibfnamefont{J.}~\bibnamefont{Yoo}}, \bibinfo{journal}{Class.
  Quant. Grav.} \textbf{\bibinfo{volume}{31}}, \bibinfo{pages}{234001}
  (\bibinfo{year}{2014}), \eprint{1409.3223}.

\bibitem[{\citenamefont{Challinor and Lewis}(2011)}]{Challinor:2011bk}
\bibinfo{author}{\bibfnamefont{A.}~\bibnamefont{Challinor}} \bibnamefont{and}
  \bibinfo{author}{\bibfnamefont{A.}~\bibnamefont{Lewis}},
  \bibinfo{journal}{Phys. Rev. D} \textbf{\bibinfo{volume}{84}},
  \bibinfo{pages}{043516} (\bibinfo{year}{2011}), \eprint{1105.5292}.

\bibitem[{\citenamefont{Bonvin et~al.}(2006)\citenamefont{Bonvin, Durrer, and
  Gasparini}}]{Bonvin:2005ps}
\bibinfo{author}{\bibfnamefont{C.}~\bibnamefont{Bonvin}},
  \bibinfo{author}{\bibfnamefont{R.}~\bibnamefont{Durrer}}, \bibnamefont{and}
  \bibinfo{author}{\bibfnamefont{M.~A.} \bibnamefont{Gasparini}},
  \bibinfo{journal}{Phys. Rev. D} \textbf{\bibinfo{volume}{73}},
  \bibinfo{pages}{023523} (\bibinfo{year}{2006}), \bibinfo{note}{[Erratum:
  Phys.Rev.D 85, 029901 (2012)]}, \eprint{astro-ph/0511183}.

\bibitem[{\citenamefont{Yoo}(2010)}]{Yoo:2010ni}
\bibinfo{author}{\bibfnamefont{J.}~\bibnamefont{Yoo}}, \bibinfo{journal}{Phys.
  Rev. D} \textbf{\bibinfo{volume}{82}}, \bibinfo{pages}{083508}
  (\bibinfo{year}{2010}), \eprint{1009.3021}.

\bibitem[{\citenamefont{Yoo et~al.}(2012)\citenamefont{Yoo, Hamaus, Seljak, and
  Zaldarriaga}}]{Yoo:2012se}
\bibinfo{author}{\bibfnamefont{J.}~\bibnamefont{Yoo}},
  \bibinfo{author}{\bibfnamefont{N.}~\bibnamefont{Hamaus}},
  \bibinfo{author}{\bibfnamefont{U.}~\bibnamefont{Seljak}}, \bibnamefont{and}
  \bibinfo{author}{\bibfnamefont{M.}~\bibnamefont{Zaldarriaga}},
  \bibinfo{journal}{Phys. Rev. D} \textbf{\bibinfo{volume}{86}},
  \bibinfo{pages}{063514} (\bibinfo{year}{2012}), \eprint{1206.5809}.

\bibitem[{\citenamefont{Montanari and Durrer}(2015)}]{Montanari:2015rga}
\bibinfo{author}{\bibfnamefont{F.}~\bibnamefont{Montanari}} \bibnamefont{and}
  \bibinfo{author}{\bibfnamefont{R.}~\bibnamefont{Durrer}},
  \bibinfo{journal}{JCAP} \textbf{\bibinfo{volume}{10}}, \bibinfo{pages}{070}
  (\bibinfo{year}{2015}), \eprint{1506.01369}.

\bibitem[{\citenamefont{Gao et~al.}(2023)\citenamefont{Gao, Raccanelli, and
  Vlah}}]{Gao:2023rmo}
\bibinfo{author}{\bibfnamefont{Z.}~\bibnamefont{Gao}},
  \bibinfo{author}{\bibfnamefont{A.}~\bibnamefont{Raccanelli}},
  \bibnamefont{and} \bibinfo{author}{\bibfnamefont{Z.}~\bibnamefont{Vlah}},
  \bibinfo{journal}{Phys. Rev. D} \textbf{\bibinfo{volume}{108}},
  \bibinfo{pages}{043503} (\bibinfo{year}{2023}), \eprint{2306.02993}.

\bibitem[{\citenamefont{Kaiser}(1987)}]{Kaiser1987}
\bibinfo{author}{\bibfnamefont{N.}~\bibnamefont{Kaiser}},
  \bibinfo{journal}{Mon. Not. Roy. Astron. Soc.}
  \textbf{\bibinfo{volume}{227}}, \bibinfo{pages}{1} (\bibinfo{year}{1987}).

\bibitem[{\citenamefont{Hamilton}(1997)}]{Hamilton1997}
\bibinfo{author}{\bibfnamefont{A.~J.~S.} \bibnamefont{Hamilton}}, in
  \emph{\bibinfo{booktitle}{{Ringberg Workshop on Large Scale Structure}}}
  (\bibinfo{year}{1997}), \eprint{astro-ph/9708102}.

\bibitem[{\citenamefont{Fisher et~al.}(1994)\citenamefont{Fisher, Scharf, and
  Lahav}}]{Fisher:1993pz}
\bibinfo{author}{\bibfnamefont{K.~B.} \bibnamefont{Fisher}},
  \bibinfo{author}{\bibfnamefont{C.~A.} \bibnamefont{Scharf}},
  \bibnamefont{and} \bibinfo{author}{\bibfnamefont{O.}~\bibnamefont{Lahav}},
  \bibinfo{journal}{Mon. Not. Roy. Astron. Soc.}
  \textbf{\bibinfo{volume}{266}}, \bibinfo{pages}{219} (\bibinfo{year}{1994}),
  \eprint{astro-ph/9309027}.

\bibitem[{\citenamefont{Cole et~al.}(1994)\citenamefont{Cole, Fisher, and
  Weinberg}}]{Cole:1993kh}
\bibinfo{author}{\bibfnamefont{S.}~\bibnamefont{Cole}},
  \bibinfo{author}{\bibfnamefont{K.~B.} \bibnamefont{Fisher}},
  \bibnamefont{and} \bibinfo{author}{\bibfnamefont{D.~H.}
  \bibnamefont{Weinberg}}, \bibinfo{journal}{Mon. Not. Roy. Astron. Soc.}
  \textbf{\bibinfo{volume}{267}}, \bibinfo{pages}{785} (\bibinfo{year}{1994}),
  \eprint{astro-ph/9308003}.

\bibitem[{\citenamefont{Linder}(2005)}]{Linder2005}
\bibinfo{author}{\bibfnamefont{E.~V.} \bibnamefont{Linder}},
  \bibinfo{journal}{Phys. Rev. D} \textbf{\bibinfo{volume}{72}},
  \bibinfo{pages}{043529} (\bibinfo{year}{2005}), \eprint{astro-ph/0507263}.

\bibitem[{\citenamefont{Kaiser}(1984)}]{Kaiser:1984sw}
\bibinfo{author}{\bibfnamefont{N.}~\bibnamefont{Kaiser}},
  \bibinfo{journal}{Astrophys. J. Lett.} \textbf{\bibinfo{volume}{284}},
  \bibinfo{pages}{L9} (\bibinfo{year}{1984}).

\bibitem[{\citenamefont{Bardeen et~al.}(1986)\citenamefont{Bardeen, Bond,
  Kaiser, and Szalay}}]{Bardeen:1985tr}
\bibinfo{author}{\bibfnamefont{J.~M.} \bibnamefont{Bardeen}},
  \bibinfo{author}{\bibfnamefont{J.~R.} \bibnamefont{Bond}},
  \bibinfo{author}{\bibfnamefont{N.}~\bibnamefont{Kaiser}}, \bibnamefont{and}
  \bibinfo{author}{\bibfnamefont{A.~S.} \bibnamefont{Szalay}},
  \bibinfo{journal}{Astrophys. J.} \textbf{\bibinfo{volume}{304}},
  \bibinfo{pages}{15} (\bibinfo{year}{1986}).

\bibitem[{\citenamefont{Mo and White}(1996)}]{Mo:1995cs}
\bibinfo{author}{\bibfnamefont{H.~J.} \bibnamefont{Mo}} \bibnamefont{and}
  \bibinfo{author}{\bibfnamefont{S.~D.~M.} \bibnamefont{White}},
  \bibinfo{journal}{Mon. Not. Roy. Astron. Soc.}
  \textbf{\bibinfo{volume}{282}}, \bibinfo{pages}{347} (\bibinfo{year}{1996}),
  \eprint{astro-ph/9512127}.

\bibitem[{\citenamefont{Desjacques et~al.}(2018)\citenamefont{Desjacques,
  Jeong, and Schmidt}}]{Desjacques:2016bnm}
\bibinfo{author}{\bibfnamefont{V.}~\bibnamefont{Desjacques}},
  \bibinfo{author}{\bibfnamefont{D.}~\bibnamefont{Jeong}}, \bibnamefont{and}
  \bibinfo{author}{\bibfnamefont{F.}~\bibnamefont{Schmidt}},
  \bibinfo{journal}{Phys. Rept.} \textbf{\bibinfo{volume}{733}},
  \bibinfo{pages}{1} (\bibinfo{year}{2018}), \eprint{1611.09787}.

\bibitem[{\citenamefont{Abramo and Bertacca}(2017)}]{Abramo:2017xnp}
\bibinfo{author}{\bibfnamefont{L.~R.} \bibnamefont{Abramo}} \bibnamefont{and}
  \bibinfo{author}{\bibfnamefont{D.}~\bibnamefont{Bertacca}},
  \bibinfo{journal}{Phys. Rev. D} \textbf{\bibinfo{volume}{96}},
  \bibinfo{pages}{123535} (\bibinfo{year}{2017}), \eprint{1706.01834}.

\bibitem[{\citenamefont{Andrianomena et~al.}(2019)\citenamefont{Andrianomena,
  Bonvin, Bacon, Bull, Clarkson, Maartens, and Moloi}}]{Andrianomena:2018aad}
\bibinfo{author}{\bibfnamefont{S.}~\bibnamefont{Andrianomena}},
  \bibinfo{author}{\bibfnamefont{C.}~\bibnamefont{Bonvin}},
  \bibinfo{author}{\bibfnamefont{D.}~\bibnamefont{Bacon}},
  \bibinfo{author}{\bibfnamefont{P.}~\bibnamefont{Bull}},
  \bibinfo{author}{\bibfnamefont{C.}~\bibnamefont{Clarkson}},
  \bibinfo{author}{\bibfnamefont{R.}~\bibnamefont{Maartens}}, \bibnamefont{and}
  \bibinfo{author}{\bibfnamefont{T.}~\bibnamefont{Moloi}},
  \bibinfo{journal}{Mon. Not. Roy. Astron. Soc.}
  \textbf{\bibinfo{volume}{488}}, \bibinfo{pages}{3759} (\bibinfo{year}{2019}),
  \eprint{1810.12793}.

\bibitem[{\citenamefont{Semenzato et~al.}(2024)\citenamefont{Semenzato,
  Bertacca, and Raccanelli}}]{Semenzato:2024rlc}
\bibinfo{author}{\bibfnamefont{F.}~\bibnamefont{Semenzato}},
  \bibinfo{author}{\bibfnamefont{D.}~\bibnamefont{Bertacca}}, \bibnamefont{and}
  \bibinfo{author}{\bibfnamefont{A.}~\bibnamefont{Raccanelli}}
  (\bibinfo{year}{2024}), \eprint{2406.09545}.

\bibitem[{\citenamefont{Salopek and Bond}(1991)}]{Salopek1990}
\bibinfo{author}{\bibfnamefont{D.~S.} \bibnamefont{Salopek}} \bibnamefont{and}
  \bibinfo{author}{\bibfnamefont{J.~R.} \bibnamefont{Bond}},
  \bibinfo{journal}{Phys. Rev. D} \textbf{\bibinfo{volume}{43}},
  \bibinfo{pages}{1005} (\bibinfo{year}{1991}).

\bibitem[{\citenamefont{Komatsu and Spergel}(2001)}]{Komatsu2001}
\bibinfo{author}{\bibfnamefont{E.}~\bibnamefont{Komatsu}} \bibnamefont{and}
  \bibinfo{author}{\bibfnamefont{D.~N.} \bibnamefont{Spergel}},
  \bibinfo{journal}{Phys. Rev. D} \textbf{\bibinfo{volume}{63}},
  \bibinfo{pages}{063002} (\bibinfo{year}{2001}), \eprint{astro-ph/0005036}.

\bibitem[{\citenamefont{Verde et~al.}(2001)\citenamefont{Verde, Jimenez,
  Kamionkowski, and Matarrese}}]{Verde:2000vr}
\bibinfo{author}{\bibfnamefont{L.}~\bibnamefont{Verde}},
  \bibinfo{author}{\bibfnamefont{R.}~\bibnamefont{Jimenez}},
  \bibinfo{author}{\bibfnamefont{M.}~\bibnamefont{Kamionkowski}},
  \bibnamefont{and}
  \bibinfo{author}{\bibfnamefont{S.}~\bibnamefont{Matarrese}},
  \bibinfo{journal}{Mon. Not. Roy. Astron. Soc.}
  \textbf{\bibinfo{volume}{325}}, \bibinfo{pages}{412} (\bibinfo{year}{2001}),
  \eprint{astro-ph/0011180}.

\bibitem[{\citenamefont{Matarrese and Verde}(2008)}]{Matarrese2008}
\bibinfo{author}{\bibfnamefont{S.}~\bibnamefont{Matarrese}} \bibnamefont{and}
  \bibinfo{author}{\bibfnamefont{L.}~\bibnamefont{Verde}},
  \bibinfo{journal}{Astrophys. J. Lett.} \textbf{\bibinfo{volume}{677}},
  \bibinfo{pages}{L77} (\bibinfo{year}{2008}), \eprint{0801.4826}.

\bibitem[{\citenamefont{Matarrese et~al.}(2000)\citenamefont{Matarrese, Verde,
  and Jimenez}}]{Matarrese:2000iz}
\bibinfo{author}{\bibfnamefont{S.}~\bibnamefont{Matarrese}},
  \bibinfo{author}{\bibfnamefont{L.}~\bibnamefont{Verde}}, \bibnamefont{and}
  \bibinfo{author}{\bibfnamefont{R.}~\bibnamefont{Jimenez}},
  \bibinfo{journal}{Astrophys. J.} \textbf{\bibinfo{volume}{541}},
  \bibinfo{pages}{10} (\bibinfo{year}{2000}), \eprint{astro-ph/0001366}.

\bibitem[{\citenamefont{Dalal et~al.}(2008)\citenamefont{Dalal, Dore, Huterer,
  and Shirokov}}]{Dalal2007}
\bibinfo{author}{\bibfnamefont{N.}~\bibnamefont{Dalal}},
  \bibinfo{author}{\bibfnamefont{O.}~\bibnamefont{Dore}},
  \bibinfo{author}{\bibfnamefont{D.}~\bibnamefont{Huterer}}, \bibnamefont{and}
  \bibinfo{author}{\bibfnamefont{A.}~\bibnamefont{Shirokov}},
  \bibinfo{journal}{Phys. Rev. D} \textbf{\bibinfo{volume}{77}},
  \bibinfo{pages}{123514} (\bibinfo{year}{2008}), \eprint{0710.4560}.

\bibitem[{\citenamefont{Barreira}(2020)}]{Barreira:2020ekm}
\bibinfo{author}{\bibfnamefont{A.}~\bibnamefont{Barreira}},
  \bibinfo{journal}{JCAP} \textbf{\bibinfo{volume}{12}}, \bibinfo{pages}{031}
  (\bibinfo{year}{2020}), \eprint{2009.06622}.

\bibitem[{\citenamefont{Barreira}(2022)}]{Barreira:2022sey}
\bibinfo{author}{\bibfnamefont{A.}~\bibnamefont{Barreira}},
  \bibinfo{journal}{JCAP} \textbf{\bibinfo{volume}{11}}, \bibinfo{pages}{013}
  (\bibinfo{year}{2022}), \eprint{2205.05673}.

\bibitem[{\citenamefont{Fondi et~al.}(2023)\citenamefont{Fondi, Verde,
  Villaescusa-Navarro, Baldi, Coulton, Jung, Karagiannis, Liguori, Ravenni, and
  Wandelt}}]{Fondi:2023egm}
\bibinfo{author}{\bibfnamefont{E.}~\bibnamefont{Fondi}},
  \bibinfo{author}{\bibfnamefont{L.}~\bibnamefont{Verde}},
  \bibinfo{author}{\bibfnamefont{F.}~\bibnamefont{Villaescusa-Navarro}},
  \bibinfo{author}{\bibfnamefont{M.}~\bibnamefont{Baldi}},
  \bibinfo{author}{\bibfnamefont{W.~R.} \bibnamefont{Coulton}},
  \bibinfo{author}{\bibfnamefont{G.}~\bibnamefont{Jung}},
  \bibinfo{author}{\bibfnamefont{D.}~\bibnamefont{Karagiannis}},
  \bibinfo{author}{\bibfnamefont{M.}~\bibnamefont{Liguori}},
  \bibinfo{author}{\bibfnamefont{A.}~\bibnamefont{Ravenni}}, \bibnamefont{and}
  \bibinfo{author}{\bibfnamefont{B.~D.} \bibnamefont{Wandelt}}
  (\bibinfo{year}{2023}), \eprint{2311.10088}.

\bibitem[{\citenamefont{Reid et~al.}(2010)\citenamefont{Reid, Verde, Dolag,
  Matarrese, and Moscardini}}]{reid2010non}
\bibinfo{author}{\bibfnamefont{B.~A.} \bibnamefont{Reid}},
  \bibinfo{author}{\bibfnamefont{L.}~\bibnamefont{Verde}},
  \bibinfo{author}{\bibfnamefont{K.}~\bibnamefont{Dolag}},
  \bibinfo{author}{\bibfnamefont{S.}~\bibnamefont{Matarrese}},
  \bibnamefont{and}
  \bibinfo{author}{\bibfnamefont{L.}~\bibnamefont{Moscardini}},
  \bibinfo{journal}{Journal of Cosmology and Astroparticle Physics}
  \textbf{\bibinfo{volume}{2010}}, \bibinfo{pages}{013} (\bibinfo{year}{2010}).

\bibitem[{\citenamefont{Sheth et~al.}(2001)\citenamefont{Sheth, Mo, and
  Tormen}}]{Sheth:1999su}
\bibinfo{author}{\bibfnamefont{R.~K.} \bibnamefont{Sheth}},
  \bibinfo{author}{\bibfnamefont{H.~J.} \bibnamefont{Mo}}, \bibnamefont{and}
  \bibinfo{author}{\bibfnamefont{G.}~\bibnamefont{Tormen}},
  \bibinfo{journal}{Mon. Not. Roy. Astron. Soc.}
  \textbf{\bibinfo{volume}{323}}, \bibinfo{pages}{1} (\bibinfo{year}{2001}),
  \eprint{astro-ph/9907024}.

\bibitem[{\citenamefont{Giannantonio et~al.}(2012)\citenamefont{Giannantonio,
  Porciani, Carron, Amara, and Pillepich}}]{Giannantonio2011}
\bibinfo{author}{\bibfnamefont{T.}~\bibnamefont{Giannantonio}},
  \bibinfo{author}{\bibfnamefont{C.}~\bibnamefont{Porciani}},
  \bibinfo{author}{\bibfnamefont{J.}~\bibnamefont{Carron}},
  \bibinfo{author}{\bibfnamefont{A.}~\bibnamefont{Amara}}, \bibnamefont{and}
  \bibinfo{author}{\bibfnamefont{A.}~\bibnamefont{Pillepich}},
  \bibinfo{journal}{Mon. Not. Roy. Astron. Soc.}
  \textbf{\bibinfo{volume}{422}}, \bibinfo{pages}{2854} (\bibinfo{year}{2012}),
  \eprint{1109.0958}.

\bibitem[{\citenamefont{Babich et~al.}(2004)\citenamefont{Babich, Creminelli,
  and Zaldarriaga}}]{Babich:2004gb}
\bibinfo{author}{\bibfnamefont{D.}~\bibnamefont{Babich}},
  \bibinfo{author}{\bibfnamefont{P.}~\bibnamefont{Creminelli}},
  \bibnamefont{and}
  \bibinfo{author}{\bibfnamefont{M.}~\bibnamefont{Zaldarriaga}},
  \bibinfo{journal}{JCAP} \textbf{\bibinfo{volume}{08}}, \bibinfo{pages}{009}
  (\bibinfo{year}{2004}), \eprint{astro-ph/0405356}.

\bibitem[{\citenamefont{Bautista et~al.}(2020)}]{Bautista:2020ahg}
\bibinfo{author}{\bibfnamefont{J.~E.} \bibnamefont{Bautista}}
  \bibnamefont{et~al.}, \bibinfo{journal}{Mon. Not. Roy. Astron. Soc.}
  \textbf{\bibinfo{volume}{500}}, \bibinfo{pages}{736} (\bibinfo{year}{2020}),
  \eprint{2007.08993}.

\bibitem[{\citenamefont{Sanchez et~al.}(2013)}]{Sanchez:2013uxa}
\bibinfo{author}{\bibfnamefont{A.~G.} \bibnamefont{Sanchez}}
  \bibnamefont{et~al.}, \bibinfo{journal}{Mon. Not. Roy. Astron. Soc.}
  \textbf{\bibinfo{volume}{433}}, \bibinfo{pages}{1202} (\bibinfo{year}{2013}),
  \eprint{1303.4396}.

\bibitem[{\citenamefont{Samushia et~al.}(2012)\citenamefont{Samushia, Percival,
  and Raccanelli}}]{Samushia:2011cs}
\bibinfo{author}{\bibfnamefont{L.}~\bibnamefont{Samushia}},
  \bibinfo{author}{\bibfnamefont{W.~J.} \bibnamefont{Percival}},
  \bibnamefont{and}
  \bibinfo{author}{\bibfnamefont{A.}~\bibnamefont{Raccanelli}},
  \bibinfo{journal}{Mon. Not. Roy. Astron. Soc.}
  \textbf{\bibinfo{volume}{420}}, \bibinfo{pages}{2102} (\bibinfo{year}{2012}),
  \eprint{1102.1014}.

\bibitem[{\citenamefont{Icaza-Lizaola et~al.}(2020)}]{eBOSS:2019gbd}
\bibinfo{author}{\bibfnamefont{M.}~\bibnamefont{Icaza-Lizaola}}
  \bibnamefont{et~al.} (\bibinfo{collaboration}{eBOSS}), \bibinfo{journal}{Mon.
  Not. Roy. Astron. Soc.} \textbf{\bibinfo{volume}{492}}, \bibinfo{pages}{4189}
  (\bibinfo{year}{2020}), \eprint{1909.07742}.

\bibitem[{\citenamefont{Tamone et~al.}(2020)}]{eBOSS:2020qek}
\bibinfo{author}{\bibfnamefont{A.}~\bibnamefont{Tamone}} \bibnamefont{et~al.}
  (\bibinfo{collaboration}{eBOSS}), \bibinfo{journal}{Mon. Not. Roy. Astron.
  Soc.} \textbf{\bibinfo{volume}{499}}, \bibinfo{pages}{5527}
  (\bibinfo{year}{2020}), \eprint{2007.09009}.

\bibitem[{\citenamefont{Szapudi}(2004)}]{Szapudi:2004gh}
\bibinfo{author}{\bibfnamefont{I.}~\bibnamefont{Szapudi}},
  \bibinfo{journal}{Astrophys. J.} \textbf{\bibinfo{volume}{614}},
  \bibinfo{pages}{51} (\bibinfo{year}{2004}), \eprint{astro-ph/0404477}.

\bibitem[{\citenamefont{Desjacques et~al.}(2020)\citenamefont{Desjacques,
  Ginat, and Reischke}}]{Desjacques:2020zue}
\bibinfo{author}{\bibfnamefont{V.}~\bibnamefont{Desjacques}},
  \bibinfo{author}{\bibfnamefont{Y.~B.} \bibnamefont{Ginat}}, \bibnamefont{and}
  \bibinfo{author}{\bibfnamefont{R.}~\bibnamefont{Reischke}}
  (\bibinfo{year}{2020}), \eprint{2009.02036}.

\bibitem[{\citenamefont{Yoo and Seljak}(2015)}]{Yoo:2013zga}
\bibinfo{author}{\bibfnamefont{J.}~\bibnamefont{Yoo}} \bibnamefont{and}
  \bibinfo{author}{\bibfnamefont{U.}~\bibnamefont{Seljak}},
  \bibinfo{journal}{Mon. Not. Roy. Astron. Soc.}
  \textbf{\bibinfo{volume}{447}}, \bibinfo{pages}{1789} (\bibinfo{year}{2015}),
  \eprint{1308.1093}.

\bibitem[{\citenamefont{Vogeley and Szalay}(1996)}]{Vogeley:1996xu}
\bibinfo{author}{\bibfnamefont{M.~S.} \bibnamefont{Vogeley}} \bibnamefont{and}
  \bibinfo{author}{\bibfnamefont{A.~S.} \bibnamefont{Szalay}},
  \bibinfo{journal}{Astrophys. J.} \textbf{\bibinfo{volume}{465}},
  \bibinfo{pages}{34} (\bibinfo{year}{1996}), \eprint{astro-ph/9601185}.

\bibitem[{\citenamefont{Tegmark}(1997)}]{Tegmark:1997rp}
\bibinfo{author}{\bibfnamefont{M.}~\bibnamefont{Tegmark}},
  \bibinfo{journal}{Phys. Rev. Lett.} \textbf{\bibinfo{volume}{79}},
  \bibinfo{pages}{3806} (\bibinfo{year}{1997}), \eprint{astro-ph/9706198}.

\bibitem[{\citenamefont{Aghanim et~al.}(2020)}]{Planck2018}
\bibinfo{author}{\bibfnamefont{N.}~\bibnamefont{Aghanim}} \bibnamefont{et~al.}
  (\bibinfo{collaboration}{Planck}), \bibinfo{journal}{Astron. Astrophys.}
  \textbf{\bibinfo{volume}{641}}, \bibinfo{pages}{A6} (\bibinfo{year}{2020}),
  \bibinfo{note}{[Erratum: Astron.Astrophys. 652, C4 (2021)]},
  \eprint{1807.06209}.

\bibitem[{\citenamefont{Bernal et~al.}(2020)\citenamefont{Bernal, Bellomo,
  Raccanelli, and Verde}}]{Bernal:2020pwq}
\bibinfo{author}{\bibfnamefont{J.~L.} \bibnamefont{Bernal}},
  \bibinfo{author}{\bibfnamefont{N.}~\bibnamefont{Bellomo}},
  \bibinfo{author}{\bibfnamefont{A.}~\bibnamefont{Raccanelli}},
  \bibnamefont{and} \bibinfo{author}{\bibfnamefont{L.}~\bibnamefont{Verde}},
  \bibinfo{journal}{JCAP} \textbf{\bibinfo{volume}{10}}, \bibinfo{pages}{017}
  (\bibinfo{year}{2020}), \eprint{2005.09666}.

\bibitem[{\citenamefont{Raccanelli et~al.}(2019)\citenamefont{Raccanelli,
  Verde, and Villaescusa-Navarro}}]{Raccanelli:2017kht}
\bibinfo{author}{\bibfnamefont{A.}~\bibnamefont{Raccanelli}},
  \bibinfo{author}{\bibfnamefont{L.}~\bibnamefont{Verde}}, \bibnamefont{and}
  \bibinfo{author}{\bibfnamefont{F.}~\bibnamefont{Villaescusa-Navarro}},
  \bibinfo{journal}{Mon. Not. Roy. Astron. Soc.}
  \textbf{\bibinfo{volume}{483}}, \bibinfo{pages}{734} (\bibinfo{year}{2019}),
  \eprint{1704.07837}.

\bibitem[{\citenamefont{Bonvin et~al.}(2016)\citenamefont{Bonvin, Hui, and
  Gaztanaga}}]{Bonvin:2015kuc}
\bibinfo{author}{\bibfnamefont{C.}~\bibnamefont{Bonvin}},
  \bibinfo{author}{\bibfnamefont{L.}~\bibnamefont{Hui}}, \bibnamefont{and}
  \bibinfo{author}{\bibfnamefont{E.}~\bibnamefont{Gaztanaga}},
  \bibinfo{journal}{JCAP} \textbf{\bibinfo{volume}{08}}, \bibinfo{pages}{021}
  (\bibinfo{year}{2016}), \eprint{1512.03566}.

\bibitem[{\citenamefont{Tansella et~al.}(2018)\citenamefont{Tansella,
  Jelic-Cizmek, Bonvin, and Durrer}}]{Tansella:2018sld}
\bibinfo{author}{\bibfnamefont{V.}~\bibnamefont{Tansella}},
  \bibinfo{author}{\bibfnamefont{G.}~\bibnamefont{Jelic-Cizmek}},
  \bibinfo{author}{\bibfnamefont{C.}~\bibnamefont{Bonvin}}, \bibnamefont{and}
  \bibinfo{author}{\bibfnamefont{R.}~\bibnamefont{Durrer}},
  \bibinfo{journal}{JCAP} \textbf{\bibinfo{volume}{10}}, \bibinfo{pages}{032}
  (\bibinfo{year}{2018}), \eprint{1806.11090}.

\bibitem[{\citenamefont{Alvarez et~al.}(2014)}]{Alvarez:2014vva}
\bibinfo{author}{\bibfnamefont{M.}~\bibnamefont{Alvarez}} \bibnamefont{et~al.}
  (\bibinfo{year}{2014}), \eprint{1412.4671}.

\bibitem[{\citenamefont{Dvali et~al.}(2000)\citenamefont{Dvali, Gabadadze, and
  Porrati}}]{Dvali:2000hr}
\bibinfo{author}{\bibfnamefont{G.~R.} \bibnamefont{Dvali}},
  \bibinfo{author}{\bibfnamefont{G.}~\bibnamefont{Gabadadze}},
  \bibnamefont{and} \bibinfo{author}{\bibfnamefont{M.}~\bibnamefont{Porrati}},
  \bibinfo{journal}{Phys. Lett. B} \textbf{\bibinfo{volume}{485}},
  \bibinfo{pages}{208} (\bibinfo{year}{2000}), \eprint{hep-th/0005016}.

\bibitem[{\citenamefont{Schmidt}(2009)}]{Schmidt:2009sv}
\bibinfo{author}{\bibfnamefont{F.}~\bibnamefont{Schmidt}},
  \bibinfo{journal}{Phys. Rev. D} \textbf{\bibinfo{volume}{80}},
  \bibinfo{pages}{123003} (\bibinfo{year}{2009}), \eprint{0910.0235}.

\bibitem[{\citenamefont{Luty et~al.}(2003)\citenamefont{Luty, Porrati, and
  Rattazzi}}]{Luty:2003vm}
\bibinfo{author}{\bibfnamefont{M.~A.} \bibnamefont{Luty}},
  \bibinfo{author}{\bibfnamefont{M.}~\bibnamefont{Porrati}}, \bibnamefont{and}
  \bibinfo{author}{\bibfnamefont{R.}~\bibnamefont{Rattazzi}},
  \bibinfo{journal}{JHEP} \textbf{\bibinfo{volume}{09}}, \bibinfo{pages}{029}
  (\bibinfo{year}{2003}), \eprint{hep-th/0303116}.

\bibitem[{\citenamefont{Koyama}(2007)}]{Koyama:2007za}
\bibinfo{author}{\bibfnamefont{K.}~\bibnamefont{Koyama}},
  \bibinfo{journal}{Class. Quant. Grav.} \textbf{\bibinfo{volume}{24}},
  \bibinfo{pages}{R231} (\bibinfo{year}{2007}), \eprint{0709.2399}.

\bibitem[{\citenamefont{Liu et~al.}(2021)\citenamefont{Liu, Valogiannis,
  Battaglia, and Bean}}]{Liu:2021weo}
\bibinfo{author}{\bibfnamefont{R.}~\bibnamefont{Liu}},
  \bibinfo{author}{\bibfnamefont{G.}~\bibnamefont{Valogiannis}},
  \bibinfo{author}{\bibfnamefont{N.}~\bibnamefont{Battaglia}},
  \bibnamefont{and} \bibinfo{author}{\bibfnamefont{R.}~\bibnamefont{Bean}},
  \bibinfo{journal}{Phys. Rev. D} \textbf{\bibinfo{volume}{104}},
  \bibinfo{pages}{103519} (\bibinfo{year}{2021}), \eprint{2101.08728}.

\bibitem[{\citenamefont{Bosi et~al.}(2023)\citenamefont{Bosi, Bellomo, and
  Raccanelli}}]{Bosi:2023amu}
\bibinfo{author}{\bibfnamefont{M.}~\bibnamefont{Bosi}},
  \bibinfo{author}{\bibfnamefont{N.}~\bibnamefont{Bellomo}}, \bibnamefont{and}
  \bibinfo{author}{\bibfnamefont{A.}~\bibnamefont{Raccanelli}},
  \bibinfo{journal}{JCAP} \textbf{\bibinfo{volume}{11}}, \bibinfo{pages}{086}
  (\bibinfo{year}{2023}), \eprint{2306.03031}.

\bibitem[{\citenamefont{Piga et~al.}(2023)\citenamefont{Piga, Marinucci,
  D'Amico, Pietroni, Vernizzi, and Wright}}]{Piga:2022mge}
\bibinfo{author}{\bibfnamefont{L.}~\bibnamefont{Piga}},
  \bibinfo{author}{\bibfnamefont{M.}~\bibnamefont{Marinucci}},
  \bibinfo{author}{\bibfnamefont{G.}~\bibnamefont{D'Amico}},
  \bibinfo{author}{\bibfnamefont{M.}~\bibnamefont{Pietroni}},
  \bibinfo{author}{\bibfnamefont{F.}~\bibnamefont{Vernizzi}}, \bibnamefont{and}
  \bibinfo{author}{\bibfnamefont{B.~S.} \bibnamefont{Wright}},
  \bibinfo{journal}{JCAP} \textbf{\bibinfo{volume}{04}}, \bibinfo{pages}{038}
  (\bibinfo{year}{2023}), \eprint{2211.12523}.

\end{thebibliography}

\end{document}